\documentclass[11pt,a4paper]{article}
\usepackage{graphicx}        % standard LaTeX graphics tool
\usepackage[latin1]{inputenc}
\usepackage[english]{babel}
\usepackage{amssymb}
\usepackage{setspace}
\def\smunp{\vspace{20pt}}

\newcommand{\luno}[1]{\|#1\|_{l^1_b}}

\newtheorem{theorem}{Theorem}[section]
\newtheorem{lemma}{Lemma}[section]

\newtheorem{proposition}{Proposition}[section]

\newenvironment{proof}[0]{\noindent{\bf proof:}}{\par\medskip}

\title{Tail resonances of FPU q-breathers \\ and their impact on the pathway
to equipartition
}

\author
{Tiziano Penati\footnote{{\sl corresponding author}: tiz77@yahoo.it} 
\\
{\small\sl Dipartimento di Matematica ``F.Enriques''}
\\
{\small\sl Via Saldini 50, 20133 Milano, Italy}
\\ 
Sergej Flach
\\
{\small \sl Max-Planck-Institut f\"ur Physik komplexer Systeme}
\\
{\small \sl N\"othnitzer Str. 38, 01187 Dresden, Germany}
\\
}
\begin{document}

\maketitle

\abstract{Upon initial excitation of a few normal modes the energy
distribution among all modes of a nonlinear atomic chain (the
Fermi-Pasta-Ulam model) exhibits exponential localization on large
time scales. At the same time resonant anomalies (peaks) are observed
in its weakly excited tail for long times preceding equipartition. We
observe a similar resonant tail structure also for exact time-periodic
Lyapunov orbits, coined q-breathers due to their exponential
localization in modal space.  We give a simple explanation for this
structure in terms of superharmonic resonances.  The resonance
analysis agrees very well with numerical results and has predictive
power. We extend a previously developed perturbation method, based
essentially on a Poincar\'e-Lindstedt scheme, in order to account for
these resonances, and in order to treat more general model cases,
including truncated Toda potentials. Our results give qualitative and
semiquantitative account for the superharmonic resonances of
q-breathers and natural packets.}

%---------------------------------------------------------------------------
%---------------------------------------------------------------------------
%
%                     CHAPTER 1: INTRODUCTION
%
%---------------------------------------------------------------------------
%---------------------------------------------------------------------------

\newpage

\section*{}
{\bf The Fermi-Pasta-Ulam model (FPU) 
  is a one dimensional
  nonlinear lattice whose dynamics, for small energy densities
  $\mathcal{E}$ and initially localized (in Fourier space)
  excitations, is characterized by a long time freezing of the
  harmonic energy among a subset of normal modes (natural
  packets). This metastable regime persists until, after times growing
  exponentially in $\frac1{\mathcal{E}}$, equipartition is
  reached.  The intermediate state is characterized by an
  exponential localization of the harmonic energies around the initially
  excited modes and the appearance of resonant deviations in the
  tail of the energy distribution. During the transition from metastability to
  stochasticity, redistribution of mode energies happens through
  the tail resonances.

  The system admits also exact periodic orbits - $q$-breathers (QB) -
  which are obtained by continuation of the normal modes of the
  linear system. QBs also show both exponential localization
  of mode energy distribution and resonant tail structures.
  The QB localization holds below an energy density threshold
  $\mathcal{E}^*$ which depends on the nonlinear parameter.  
  The tail resonances are due to small denominators which
  appear naturally in a Poincare-Lindstedt perturbation scheme.
  
  We observe that the dynamics of the nonlinear FPU model
  generated by initially exciting one normal mode (FPU trajectory) 
  exhibits the main elements of the nearby $q$-breather. Thus the
  presented analytical approach, besides describing the
  $q$-breather, also explains some relevant features of the
  corresponding FPU trajectory. Remarkably, this correspondence
  may hold on very long times, which requires further rigorous
  explanations.}

\smunp
\hrulefill

\section{Introduction}

%------------------------------------------------------------------------------
%HISTORY

\subsection{Brief history}

The computational experiments which Fermi, Pasta and Ulam performed
around 1950 in the Los Alamos laboratory are nowadays considered as an
important step in the progress of understanding nonlinear
dynamics. The main aim of their numerical investigation was to
observe, as they expected, a ``relatively fast'' and ``equally
distributed'' sharing of the averaged harmonic energy of an anharmonic
atomic chain among all the normal modes, even when starting with
initial conditions far from some equipartition regime. When chosing
initial data localized on just one low frequency mode they however did
not observe any actual equipartition, but a localization of the energy
among some initial modes in the nearby frequency sector (also coined
in the literature and hereafter natural packets), including some
additional almost complete recurrence to the initially excited mode.

Among the several explanations that followed, we recall the one given
in 1965 by Izrailev and Chirikov \cite{IC65,IC66b}.  
They suggested the idea that the KAM theorem was the
right approach in order to explain the regular dynamical behaviour observed: if
the energy given to the system is below a threshold $E^*(N)$, then the
dynamics is influenced by the persistence of most of the nonresonant
invariant tori. Since $\lim_{N\to\infty}{\mathcal{E}^*}=0$, where
$\mathcal{E}^* = \frac{E^*(N)}{N}$ is the energy density, the FPU
phenomenon should not be relevant in the thermodynamic limit at any
finite temperature or energy density.  In 1970, instead, Bocchieri,
Scotti, Bearzi and Loinger \cite{BSBL} conjectured the existence of a
specific energy threshold $\mathcal{E}^*$ (which remains positive in
the thermodynamic limit) below which equipartition is not
reachable. Recent numerical results \cite{BeGG,BeGP,PP04} indicate
that below that threshold equipartition is still reached for infinite
times, yet the time scale of the route to equipartition may diverge
exponentially with lowering the energy density.

Several computational experiments have been performed
(e.g. \cite{DLR95,DelLL95,UllLC00,BeGG,BeGP,GPP}), suggesting possible
answers about the time needed to approach equipartition, its scaling
with the energy density and dimension of the system and the features
of the dynamics during this transition.  Only few analytical studies
touch the case of large systems: here we mention some relevant results
related to the application of the KAM Theorem
\cite{MR1831098,RINKnish}, to some special (periodic) solutions
\cite{MR1464250,RINKinv} and to a normal form construction via a PDE
\cite{BPon05}.

Recently Flach, Ivanchenko and Kanakov \cite{FlaIvaKan05} have
focussed on the main FPU observation that the initially excited normal
mode shares its energy for long times only with a few other modes from
a frequency neighbourhood in modal space.  They have identified this
long lasting regime as a dynamical localization effect and applied the
methods developed for discrete breathers in FPU chains
\cite{MR2133463} to the dynamics of normal modes. The result is that
q-breathers (QB) - time-periodic and modal-space-localized orbits -
persist in the FPU model.  The dynamics generated by one initially
excited mode evolves close to the related q-breathers for very long
times. Thus many features of the short- and medium-time evolution of
natural packets are encoded in the profile of a QB and the phase space
flow of small fluctuations around a given QB.

%--------------------------------------------------------------------------
%MODEL

\subsection{The FPU model}

Fermi, Pasta and Ulam considered a non-linear one-dimensional lattice,
with fixed boundary conditions
\begin{equation}
\label{1.hampart}
\begin{cases}
{
  H(\underline{x},\underline{y})={\frac{1}{2}}\sum_{j=1}^N{y_j^2}+\sum_{j=0}^N{V(x_{j+1}-x_j)},\cr
  x_0 = x_{N+1} = 0, 
}
\end{cases}
\end{equation}
where $V(r)=\frac12 r^2+\frac{\alpha}{3}r^3+\frac{\beta}{4}r^4$ is the
coupling potential, $(x_j,y_j)$ are respectively the displacement from
the equilibrium of the j-particle and its kinetic momentum, while the
parameters $\alpha ,\beta$ may be considered as real and positive. The
case of linear equations of motion, given by the quadratic Hamiltonian
\begin{displaymath}
 H_0(\underline{x},\underline{y})=\frac12\sum_{j=0}^N\left[y_j^2+(x_{j+1}-x_j)^2\right],
\end{displaymath}
is integrable and corresponds to a system of $N$ uncoupled normal mode
oscillators.  We can show this via the change of coordinates
\begin{eqnarray*}
Q_l &=& \sqrt{\frac{2}{N+1}}\sum_{j=1}^{N}{x_j\sin\left({\frac{jl\pi}{N+1}}\right)}\\
P_l &=& \sqrt{\frac{2}{N+1}}\sum_{j=1}^{N}{y_j\sin\left({\frac{jl\pi}{N+1}}\right)}\\
\end{eqnarray*}
which diagonalizes the Hamiltonian $H_0$ 
\begin{displaymath}
H_0(\underline{Q},\underline{p})={\frac{1}{2}}\sum_{l=1}^{N}{(P_j^2+\omega_j^2Q_j^2)},
\end{displaymath}
where $(Q_j,p_j)$ are the normal mode coordinates and momenta, and $\omega_j$ are
the normal mode frequencies given by
\begin{displaymath}
\omega_j=2\sin\left(\frac{j\pi}{2N+2}\right).
\end{displaymath}
The effect of the non-linear part in the model (\ref{1.hampart}) is to introduce
an interaction between the
different normal mode oscillators. In this way they exchange energy and the harmonic
actions are in general no longer preserved.  Even if the considered change to
normal modes appears hard to develop for the cubic and quartic part of the
Hamiltonian, there is a mathematical way to exhibit which are the {\sl
selection rules} that couple the different oscillators (see also
\cite{DelLL95}). Let us consider the so called $\alpha$-model,
where $\beta=0$. The cubic part of the potential can be
rewritten as
\begin{displaymath}
{\frac{\alpha}{3}}\sum_{j=0}^N{(x_{j+1}-x_j)^3}={\frac{\alpha}{3\sqrt{2N+2}}}\sum_{1\leq i, j, h\leq N}{\omega_i\omega_j\omega_h
  Q_iQ_jQ_h B_{ijh}},
\end{displaymath}
where we set $ B_{ijh} = \sum_{{\sigma_{1,2}=\pm 1}}{\Delta_{i+\sigma_1 j+\sigma_2 h}}$,
with
\begin{equation}
\label{1.cubrules}
\Delta_{i\pm j \pm h} = \left\{\matrix{1 & ,i\pm j \pm h=0\cr -1 &
    ,i\pm j \pm h=2(N+1)\cr 0 & ,{\rm otherwise}}\right..
\end{equation}
The equations of motion for a normal mode oscillator then read
\begin{equation}
\label{1.motion}
\ddot{Q_j}+\omega_j^2 Q_j = -\frac{\alpha}{\sqrt{2(N+1)}}\sum_{1\leq i, h\leq N}{\omega_j\omega_i\omega_h
  Q_iQ_h B_{ijh}}.
\end{equation}

%-----------------------------------------------------------------------------
%PAKETS

\subsection{Natural packets}

\begin{figure}
\includegraphics[width=0.9\columnwidth]{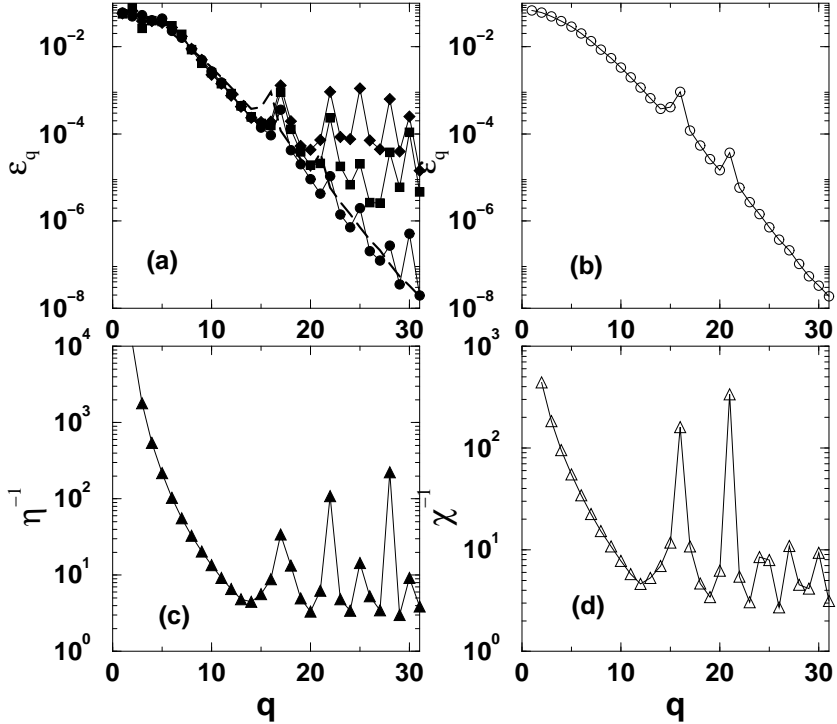}
\caption{ (a): Natural packet evolution initially exciting mode
$q_0=1$. The energies of normal modes are plotted versus $q$. Circles
- $10^4$, squares - $10^5$, rhombs - $10^6$.  Dashed line -
$q$-breather from (b) for comparison.  (b): The energies of normal
modes versus $q$ for a $q$-breather with $q_0=1$.  (c): sequence
$\eta_j^{-1}$.  (d): sequence $\chi_j^{-1}$.  Parameters in all cases:
$N=31,\,\alpha=0.33,\,\mathcal{E}=0.01$.}
\label{fig1}
\end{figure}

\begin{figure}
\includegraphics[width=0.9\columnwidth]{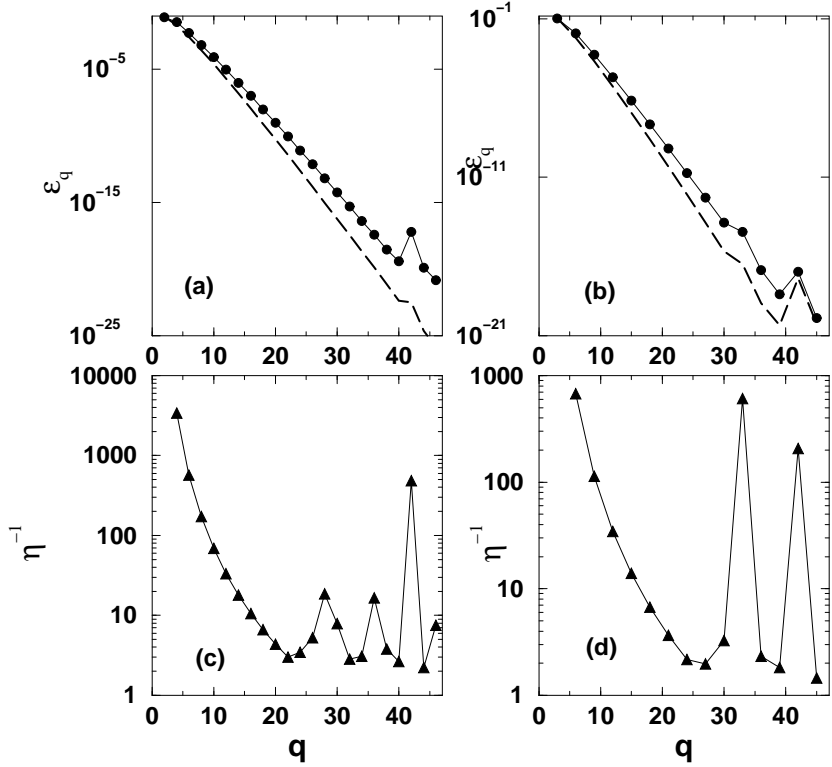}
\caption{ (a): Natural packet evolution initially exciting mode
$q_0=2$ (circles).  The energies of normal modes are plotted versus
$q$ at $t=10^5$.  The dashed line shows the energy distribution for a
corresponding $q$-breather with $q_0=2$.  (b): Same as (a), but for
$q_0=3$.  (c): sequence $\eta_j^{-1}(2,47)$.  (d): sequence
$\eta_j^{-1}(3,47)$.  Parameters in all cases:
$N=47,\,\alpha=0.1,\,\mathcal{E}=0.0025$.}
\label{fig2}
\end{figure}

One of the findings which can be related to Bocchieri's conjecture is
the so called 'metastability' of natural packets, which has been
numerically investigated by A.Giorgilli and others
\cite{BeGG,BeGP,GPP,PP04}.  What has been observed in \cite{BeGG} is
that, for energy density small enough, an initial excitation of the
first mode is spread among nearby modes up to a {\em natural
frequency} which depends only on the energy density, according to the
power law $\omega\sim\mathcal{E}^{1/4}$.  This localization of the
harmonic energy on a subset of modes, which can coexist with chaotic
dynamics inside the subset, persists for times which can grow
exponentially with the inverse of the energy density (see
\cite{BeGP}), and only after this long transient the system reaches a
proper equipartition regime.  The time-dependence of the size of these
natural packets of normal modes show two time scales $\tau_{1,2}$
which separate strongly when the energy density is below a certain
threshold. The first time scale $\tau_1$ characterizes the time needed
for the formation of a natural packet of nearly fixed (and quite
localized) shape in normal mode space. The second time scale $\tau_2$,
which becomes much larger compared to the first one when the energy
density is lowered, characterizes the time needed to finally
delocalize the packet over all modes and to reach equipartition.  Both
time scales $\tau_1 \approx \tau_2$ at some energy density threshold.
Thus at and above the energy density threshold no metastable or
intermediate normal mode localization occurs, and the natural packet
spreads more or less continuously in time over all modes. Further
numerical simulations \cite{BeGP} used the spectral entropy indicator
(which gives the effective fraction of modes involved in the dynamics)
and showed that the second time scale nicely fits with an exponential
law.

An example of the evolution of such a packet for energy densities
below the threshold (when $\tau_1 <10^4 ,\, \tau_2 >10^9$) was
observed in \cite{GPP} (Fig.3 therein). Following that work, we
performed a similar simulation, see Fig.\ref{fig1}(a). The normal mode
$q_0=1$ has been initially excited, and the harmonic energy
distribution is shown for three different times
$t=10^4,10^5,10^6$. Since harmonic mode energies may fluctuate, we
plot the time-averaged quantities, with a time-averaging window of
$10^4$. The logarithmic scale reveals that for $t=10^4$ the overall
strong exponential distribution, which is locked over many decades in
time, shows pronounced resonant peaks in the far tails (note however
that even in the peak maxima the mode energies are orders of magnitude
smaller than the energies of the main modes excited).  At $t=10^5$ a
gradual change is observed - the resonant peaks in the tail slowly
grow and drag the tail upwards. At $t=10^6$, approaching $\tau_2$,
this process continues and the slow but steady spreading of the packet
over all modes can be already anticipated. At even larger times, the
harmonic energy becomes equally distributed over all the normal modes
(not displayed here).

In Fig.~\ref{fig2}(a,b) instead, the initial excitation is localized
on the second and third mode, respectively. The profiles are shown for
time $\tau_1 < t=10^5 < \tau_2$. They are again characterized by the
existence of peaks in the tails of the distributions. The persistence
of these anomalies when changing the initial mode number excited,
suggests that there is a robust and generic pathway of equipartition
which goes via excitation of these resonant tail peaks.

Let us compare these results to the profiles of corresponding
$q$-breathers \cite{FlaIvaKan05}.  We compute periodic orbits, which
correspond to only one normal mode with $q_0$ being excited in the
limit of vanishing nonlinearity.  In Fig.\ref{fig1}(b) the profile of
a $q$-breather with $q_0=1$ is shown.  The harmonic normal mode
energies are averaged over one period of the $q$-breather.  We observe
exponential localization, and in the tail two distinct resonant
peaks. For comparison we plot the same result as a dashed line in
Fig.\ref{fig1}(a). The exponential profiles of the $q$-breather and
the corresponding natural packet are strikingly similar in the whole
$q$-space. That supports previous results that a natural packet for $
t \ll \tau_2$ evolves very close to a corresponding $q$-breather
orbit. The tail resonance locations though differ for both
trajectories. That signals that the nature of the tail resonances can
be understood already by studying a $q$-breather, but their location
is perhaps very sensitive to small perturbations. Nevertheless some
common features are also observed - the tail resonances start to
appear in the second part of the $q$-space. The dashed lines in
Fig.\ref{fig2}(a,b) are $q$-breather profiles for corresponding
smaller energies. We note that here not only the exponential decay
fits well with the natural packets, but now also the location and even
the strength of the tail resonances agrees well with the one of the
natural packets. We conclude that the correspondence between natural
packets and $q$-breathers improves with lowering the energy also with
respect to the sensitive tail resonance structures.

Further similarity between natural packets and $q$-breathers follows
from \cite{IvaKanMshFla06} where it was shown, that the localization
length of a QB depends on the expansion parameters of the potentials,
the energy density $\mathcal{E}$ and the normalized wave number ${\bf
k_0}=q_0/N$, but not anymore on the system size $N$.  In fact QBs can
be scaled to and obtained in infinitely large lattices
\cite{KanFlaIvaMsh06}.  Using estimates from perturbation theory we
can conclude that delocalization of QBs will happen when the energy
density exceeds some critical value (with the other relevant
parameters being fixed).  It can be expected then that an initially
localized excitation will quickly spread over all modes as well. This
is in contrast to the case when the energy density is below the QB
threshold. Then the QB is exponentially localized, and an initially
localized excitation is known to spread a bit and take a form similar
to a QB for long times, with subsequent departure for much longer
times. That gives further evidence that the exact and simple periodic
orbits - q-breathers - describe the metastable features of a natural
packet spreading in modal space. It is then also natural to study
whether the q-breather shows similar resonant tail peak features as
the natural packet which evolves close to the QB.  For example, in
Fig.~\ref{fig1}(b) the exact QB corresponding to $q_0=1$ shows an
exponential decay which persists also in the tail, while the tail's
harmonic distribution of the close orbit (Fig.~\ref{fig1}(a)) is more
irregular. Moreover, the QB also exhibits, even if in different sites,
a couple of resonant peaks.

In the following we will first (Section 2.) formulate a simple
superharmonic resonance analysis and show that it accounts
quantitatively for the observed tail peak positions. Then (Section 3.)
we will perform a modified perturbation approach to the q-breather
profile and provide a semiquantitative answer for the tail profiles.

%---------------------------------------------------------------------------
%---------------------------------------------------------------------------
%
%                     CHAPTER 2: TAIL RESONANCES
%
%---------------------------------------------------------------------------
%---------------------------------------------------------------------------

\section{Tail resonances}

%----------------------------------------------------------------------------
%Q-BREATHERS

\subsection{$q$-breathers}

As we noticed concluding the previous section, a fruitful way to
investigate the dynamics for very localized initial excitations is to
study the existence and the related properties of nearby invariant
manifolds.  we thus focus first on the Lyapunov orbits, called
$q$-breathers due to their localization in the normal mode
space. These orbits are the natural continuation, for small values of
a coupling parameter, of the uncoupled oscillatory solutions which
appear in the linear problem \cite{FlaIvaKan05}. The existence of
these families for every linear frequency $\omega_j$ of
(\ref{1.hampart}) goes through classical continuation theorems, such
as the Lyapunov Theorem or the Poincar\'e Theorem, and essentially
relies on the property of rational independence of any couple of
frequencies composing the spectrum. Indeed old (\cite{ConJon76}) and
more recent (\cite{DVZAN}) results on sums of roots of Unity have
proved that any linear combination of two different frequencies with
rational coefficients is always different from zero
\begin{equation}
\label{1.indfreq}
k_j\omega_j + k_l\omega_l\not=0,\qquad\qquad \forall k_j,k_l\in
\mathbb{Q},\qquad\qquad\omega_j\not=\omega_l.
\end{equation}
In order to apply the above mentioned Theorems, it is enough to use a
weaker result which ensures that (\ref{1.indfreq}) holds when $k_l=1$
and $\omega_j$ is the frequency of the linear periodic orbit we want
to continue. The new periodic orbits are expected to be continuous
deformations of the original ones upon changing the nonlinearity
parameter. Thus, if we initially excite only one $q_0$-mode, we can
imagine the dynamics being strongly influenced by the corresponding
$q_0$-breather: in fact, the initial condition is close to this
special orbit in phase space.

The idea to interpret the original results obtained by Fermi, Pasta
and Ulam in terms of deformation of periodic Lyapunov orbits has been
recently used in \cite{FlaIvaKan05}. In this paper the authors perform
both a numerical and analytical investigation of the $\alpha$ and
$\beta$ models. Their study gives an exponential localization of the
energy among the modes involved in the periodic orbit; these modes are
selected according to the definition of the coupling coefficients
$B_{ijh}$ (see eq (\ref{1.cubrules})). Furthermore, the linear
stability analysis for the $\beta$ model remarkably reveals that this
exponential localization, which gets weaker when increasing the
coupling parameter $\beta$, is still persistent when the orbit becomes
unstable. From the analytical point of view, the authors provide an
application of the century old proof of Lyapunov to the existence of
q-breathers, and a perturbative Poincar\'e-Lindstedt expansion which
explains the localization for the modes $kq_0$ belonging to the low
frequency part of the spectrum, {\em id est} when both $q_0\ll N$ and
$kq_0\ll N$. This perturbation scheme also shows how, at every step
$k$, only one new mode, precisely the mode $kq_0$, is involved; then,
by using a cubic expansion of the spectrum to cope with the small
divisors, the dynamics of this mode is approximated as an oscillation
whose frequency is $k\omega_{q_0}$.

In order to better understand the idea that moved our research, we
recall here just the main steps of their deduction, and will extend
this approach in the next section.  The starting point consists in
expanding the periodic solution with respect to a perturbation
parameter, here chosen as $\sigma=\frac{\alpha}{\sqrt{2N+2}}$
\begin{equation}
\label{2.sigmaexp}
Q_q(t) = \sum_{j\geq 0}{\sigma^j Q_q^{(j)}(t)}
\end{equation}
in (\ref{1.motion}). Sorting the terms with respect to powers of
$\sigma$, we get an infinite sequence of differential equations, one
for each component $Q_q^{(j)},\,j\geq 0$; the zero order solution is
simply the cosinusoidal oscillation of the $q_0$ mode (or the linear
approximation)
\begin{displaymath}
Q_q^{(0)} = \delta_{q,q_0}A_{q_0}\cos{(\omega_{q_0}t)}.
\end{displaymath} 
For any $n\geq 2$, the scheme then follows these rules:

\begin{itemize}
\item at every step $n-1$, only one new mode $Q_{nq_0}^{(n-1)}(t)$ is
  involved in the motion. This is related to the coupling coefficients
  $B_{nq_0,l,m}$.

\item up to the step $n-1$, every solution $Q_q^{(h)},0\leq h\leq
 n-2,q_0\leq q\leq (n-1)q_0$ is a finite sum of harmonics having
 frequencies $m\omega_{q_0},\,m\leq h+1$. Among these solutions, the
 one $Q_{jq_0}^{(j-1)}(t),\,j=2,\ldots,n-1$ are approximated with the
 largest amplitude oscillation. Therefore the new forced oscillator
 becomes
 \begin{displaymath}
   \frac{d^2}{dt^2}Q_{nq_0}^{(n-1)} + \omega^2_{nq_0}Q_{nq_0}^{(n-1)}
   = \sum_{m=0}^n{C_m\cos{(m\omega_{q_0}t)}},
 \end{displaymath}
 with a suitably defined sequence $C_m$; the initial conditions are
 chosen such that the solution reads
 \begin{equation}
   \label{2.solut}
   Q_{nq_0}^{(n-1)} = \sum_{m=0}^n{\frac{C_m}{\omega^2_{nq_0} -
   m^2\omega^2_{q_0}}\cos{(m\omega_{q_0}t)}}.
 \end{equation}
 
\item exploiting the cubic expansion at low frequencies $\omega_k
  \approx k\delta-\frac1{24}k^3\delta^3,\,\delta=\frac{\pi}{N+1}$, it
  is possible both to realize that the smallest divisor is
  $\omega^2_{nq_0} - n^2\omega^2_{q_0}$ and to give an approximated
  value of this quantity as
  \begin{equation}
  \label{2.apprlow}
   \omega^2_{nq_0} - n^2\omega^2_{q_0}\approx -\frac1{12}n^2(n^2-1)\omega_{q_0}^4.
  \end{equation} 
  Hence, the solution can still be reduced to a single oscillator
  \begin{displaymath}
    Q_{nq_0}^{(n-1)} =
    \frac{12C_n}{n^2(n^2-1)\omega_{q_0}^4}\cos{(n\omega_{q_0}t)}\left(1+\mathcal{O}\left(\frac{nq_0}{N}\right)^2\right)
  \end{displaymath}
if the error $\mathcal{O}\left(\frac{nq_0}{N}\right)^2$ is small.
\end{itemize}

%-------------------------------------------------------------------------------
%HIGHER RESONANCES

\subsection{Higher order nonlinear resonances}

\begin{figure}
\includegraphics[width=0.9\columnwidth]{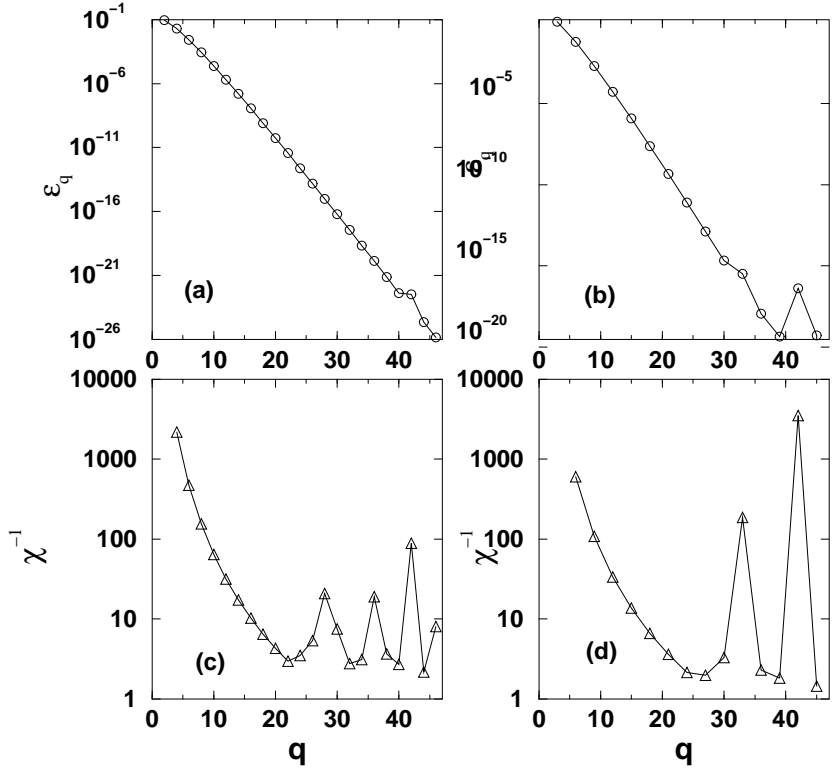}
\caption{Parameters: $N=47,\,\alpha=0.1,\,\mathcal{E}=0.0025$. (a):
Energy of normal modes versus mode number of a QB for $q_0=2$. (b):
Similar to (a), but for a QB with $q_0=3$. (c): sequence
$\chi_j^{-1}(2,47)$. (d): sequence $\chi_j^{-1}(3,47)$.}
\label{fig3} 
\end{figure}

The perturbation scheme used in \cite{FlaIvaKan05} is based on the
cubic approximation of the dependence of $\omega_j$ on $j$. This
approximation holds in the low frequency domain and allows to perform
a detailed analysis of the small divisors. In this way one limits the
study to the first third of the spectrum from below, were this
approximation holds. However, the numerical computation of these
orbits clearly shows an exponential decay which continues into the
upper two thirds of the normal mode frequency spectrum. However, some
of these orbits also show resonant deviations in the tails. We
conjecture that these resonances are due to almost resonances which
are not captured by the abovementioned cubic approximation of the
dependence of normal mode frequencies on mode numbers.  Thus we
investigate the sequence of small divisors
\begin{equation}
\label{2.eta}
\overline{\eta}_{n,m} = |\omega_{nq_0}^2 -
m^2\omega_{q_0}^2|\qquad\qquad m=n,n-2,n-4,\ldots
\end{equation}
for every frequency $\omega_{nq_0}$ belonging to a mode involved in
the $q$-breather considered. The special choice of only odd or even
multiples is related to the perturbation theory and will become clear
later. Then, for each $n$, we can also consider the sequence of the
smallest divisors
\begin{equation}
\label{2.overeta}
\eta_n = \min_{m=n,n-2,\ldots}{\{\overline\eta_{n,m}\}}.
\end{equation}
For each $q_0$, the above definitions refers to $\omega_{q_0}$, which
is the first order approximation of the true frequency $\Omega_{q_0}$
of the $q_0$-breather considered. Equations (\ref{2.eta}) and
(\ref{2.overeta}) are a consequence of applying the
Poincar\'e-Lindstedt method, since these are the small denominators
appearing in the Fourier expansion of the leading solution
(\ref{2.solut}). However the approximating $\Omega_{q_0}$ by
$\omega_{q_0}$ is less accurate the larger we take the product of
perturbation parameters $\alpha \mathcal{E}$. Thus intuition suggests
to consider also sequences obtained by replacing $\omega_{q_0}$ with
the true QB frequency $\Omega_{q_0}(\alpha,\mathcal{E})$
\begin{equation}
\overline\chi_{n,m} = |\omega_{nq_0}^2 - m^2\Omega_{q_0}^2|,\qquad\qquad \chi_n
= \min_{m=n,n-2,\ldots}{\{\overline\chi_{n,m}\}},
\end{equation}
with the same choice of $m$. These are the small denominators
appearing in a pure Fourier expansion of the solutions $Q_{nq_0}$ by
substituting $Q_{nq_0} = \sum_{k\in \mathbb{Z}}{\hat Q_{nq_0,k}
e^{ik\Omega_{q_0}t}}$ in (\ref{1.motion}).

In Fig.\ref{fig1}(c,d) we plot the sequences $\eta_n^{-1}$ and
$\chi_n^{-1}$ for $q_0=1$, $N=31$ and $\mathcal{E}=0.01$, and in
Fig.~\ref{fig2}(c,d) and Fig.\ref{fig3}(c,d) respectively the
sequences $\eta_n^{-1}$ and $\chi_n^{-1}$ for
$q_0=2,3,\,N+1=48,\,\alpha=0.1$ and $\mathcal{E}=0.0025$.  Note that
the corresponding true QB frequency $\Omega_{q_0}$ has been used to
compute $\chi$.  In all cases the first part of the sequence is
influenced by the cubic approximation. The curves follow the law
$\frac1{3\log{(k\delta)}}$ and the largest term corresponds to the
smallest allowed value of $q$.  Far from the low frequency sector, we
can observe a different and irregular behaviour.  It happens to some
of the frequencies to be strongly almost resonant with multiples of
the QB frequency $\omega_{q_0}$ (or $\Omega_{q_0})$.  Let us estimate
the lower bound $k_c q_0$ for the existence of such strong almost
resonances. It is enough to find the first $k_c$ for which
\begin{displaymath}
\omega_{k_c q_0} - (k_c-1)\omega_{q_0}=0
\end{displaymath}
holds.  Expanding $\omega_q$ for small $q$ to third order we find
$k_c\approx (\frac{6 N^2}{\pi^2 q_0^2})^{\frac13}$, or in
size-independent quantities $\kappa_c \approx
(\frac{6\kappa_0}{\pi^2})^{\frac13}$.  This value agrees well with the
location of the first local minimum in the mentioned figures.  The
subsequent resonances suggest that the QB profile will show resonant
deviations in its tail, exactly at the predicted locations.  The
sequence $\chi_j$ for very small nonlinearities can be well
approximated by $\eta_j$. Increasing the perturbation, new small
denominators may appear and some of the old one can become less
relevant due to the QB frequency change.
 
In Fig.~\ref{fig1}(d) the $\chi$ sequence corresponds to the QB in
Fig.~\ref{fig1}(b). The resonant local minima for $\chi_j$ are $j=16$
and $j=21$, which perfectly agree with the two tail peaks of the
corresponding QB. Note that the $\eta$ sequence in Fig.~\ref{fig1}(c)
shows almost the same resonance locations, but the relative strength
is different.  Also both sequences show additional resonance
locations, which are not observed in the QB tail. Instead almost all
of them are observed in the energy distribution of the corresponding
natural packet in Fig.~\ref{fig1}(a).

The distributions in Fig.~\ref{fig2}(a,b) refer to two natural packets
localized respectively on $q_0=2$ and $q_0=3$, while in
Fig.~\ref{fig3}(a,b) we show the corresponding exact
$q_0$-breathers. the profiles, including the (much weaker) tail
resonances, coincide remarkably. That is due to the weak nonlinearity.
The corresponding $\eta$ sequencies in Fig.~\ref{fig2}(c,d) and $\chi$
sequencies Fig.~\ref{fig3}(c,d) also show very good agreement with the
really observed locations of tail resonances.

A conjecture which explains the existence of the observed peaks in the
tails can be now formulated. The amplitudes of some modes are
amplified by very small denominators generated via strong almost exact
resonances of the type $\omega_{nq_0}^2-m^2\Omega_{q_0}^2$.  The
$\eta$ and $\chi$ sequencies allow to predict possible locations of
these superharmonic resonances, however not always the predicted
locations are indeed occupied by observed resonances.

Let us now formulate a set of QB properties which have to be
studied. First we recall again that the energy distribution of the
computed $q$-breathers for $q_0\ll N$ shows that their exponential
localization is kept throughout the whole $q$-space and is not
restricted to the low $q$ mode sector, as anticipated by the
approximations used in perturbation theory. Hence, it is an important
task to extend the perturbation theory developed in \cite{FlaIvaKan05}
in order to include the large $q$ sector. This would imply to deal
with the small divisors without any special Taylor
expansion. Moreover, to properly cope with the cascade of modes, one
should understand which are the actual modes excited in the
$q$-breather, according to the coefficient $B_{ijh}$.

Next, as suggested by several numerical studies, QBs appear to have
exponential localization for any $q_0$.  One can expect that this is a
common property of all the Lyapunov orbits in FPU chains. The expected
differences, when changing $q_0$, could concern the sequence of modes
involved and, maybe, details of the exponential decay and stability of
QBs.

As shown above, tail resonances in the QB energy distribution appear
due to superharmonic almost resonances. These tail resonances are also
observed in the evolution of natural packets, and are responsible for
the approach to equipartition, pumping energy from the packet core
into the tail. An analytical theory would thus allow to understand and
describe the ultimate route to equipartition, the holy grail of the
FPU problem.

Below we will address some of the above points.

%---------------------------------------------------------------------------
%---------------------------------------------------------------------------
%
%                     CHAPTER 3: PERTURBATION THEORY
%
%---------------------------------------------------------------------------
%---------------------------------------------------------------------------

\section{Improved perturbation theory}

In this third part we present a result based on the
Poincar\'e-Lindstedt (PL) perturbation scheme: for a generic
$q_0$-breather, we prove the existence of a threshold
$\mathcal{E}_{q_0}^*$, scaling with $\kappa_0=\frac{q_0}{N+1}$ and
with the perturbation parameter, such that for the energy density
$\mathcal{E}_{q_0}<\mathcal{E}_{q_0}^*$ the averaged harmonic energies
involved in the QB dynamics are exponentially localized.  In the
$\alpha$ model, for example, we get for any $n\geq 2$
\begin{eqnarray}
\label{3.firstest}
\langle E_{\overline q_n} \rangle &<& E_{q_0}
(\omega_{nq_0}^2+n^2\omega_{q_0}^2)\left(\frac{\mu_n}{\overline\mu_{n-1}}
\right)^2 e^{-(1-n)|\ln{\gamma_n}|},\cr \gamma_n &=& \mathcal{C}
(\alpha\overline\mu_{n-1})^2 \mathcal{E}_{q_0},
\end{eqnarray}
where $\overline q_n$ is the sequence of the modes excited in the QB,
$\langle E_{\overline q_n} \rangle$ is the leading averaged harmonic
energy
\begin{equation}
\label{3.avharm}
\langle E_{\overline q_n}\rangle =
\frac{\sigma^{2n-2}}{\tau}\int_0^\tau{\frac12 \left[
\omega_{nq_0}^2(Q_{{\overline q}_n}^{(n-1)})^2(t) + (P_{{\overline
q}_n}^{(n-1)})^2(t)\right]dt},
\end{equation}
and $\mu_n,\,\overline\mu_n$ are defined as
\begin{displaymath}
\mu_n := \frac1{\eta_n},\qquad\qquad\qquad
\overline{\mu}_n=\max_{l=2,\ldots,n}{\{\mu_l\}}.
\end{displaymath}
This result completes the study on QB localization in Fourier space
performed in \cite{FlaIvaKan05}, developing a more general estimate
based on the same PL scheme. Among the differences between the two
approaches, we first mention the lack of an explicit local
approximation for the smallest divisor as in (\ref{2.apprlow}): in
fact for any $n$ we simply factorize the term $\eta_n$ without knowing
its value. This allows us to generalize the previous result, once it
is clear how to determine $\overline q_n$. Another difference concerns
the harmonic cutoff of the leading solution (\ref{2.solut}): in order
to perform a rigorous approach we have decided to keep all the
harmonics in the Fourier expansion of the solution. This leads to the
constant factor $\mathcal{C}>1$ in (\ref{3.firstest}), which takes
into account both the accumulation of frequencies' products and all
the possible non-zero higher harmonics. This affects the threshold
$\mathcal{E}_{q_0}^*$, which is derived from (\ref{3.firstest}) by
requesting $\gamma_n<1$:
\begin{equation}
\label{3.thresh}
\mathcal{E}_{q_0}^*:=\mathcal{C}\left(\frac{1}{\alpha
\mu}\right)^2,\qquad\qquad\mu:=\max{\{\overline\mu_n\}}.
\end{equation}
We have already mentioned the scaling properties of
$\mathcal{E}_{q_0}^*$, which depends on the wave number $\kappa_0$,
but not on the actual system size.  The main reason is that, when
changing both $q_0$ and $N+1$ while keeping their ratio fixed, the
frequencies involved in the QB evolution are always the same. Then the
small denominators also depend only on $\kappa_0$ and the same holds
for $\mu$.  On the other hand, if we fix $q_0$ while changing the size
$N$ of the system, the smallest possible divisor ${\overline\eta}_j$
goes to zero. Consequently the threshold $\mathcal{E}_{q_0}^*$
vanishes and the QB loses its localization in such a case. In general
we are not able to provide an explicit value for $\mu$.  In some
cases, however, it is possible to make a suitable approximation
similar to (\ref{2.apprlow}). For example, when $q_0$ is close to zero
or to $N$, it follows
\begin{displaymath}
\label{3.lowhighalpha}
\mathcal{E}^*_{q_0}\approx \alpha^{-2}\pi^4 \kappa_0^4,\qquad\qquad
\mathcal{E}^*_{q_0}\approx \alpha^{-2}\pi^4 (1-\kappa_0)^4,
\end{displaymath}
for the $\alpha$ model and
\begin{displaymath}
\label{3.lowhighbeta}
\mathcal{E}^*_{q_0}\approx \beta^{-1}\pi^2 \kappa_0^2,\qquad\qquad
\mathcal{E}^*_{q_0}\approx \beta^{-1}\pi^2 (1-\kappa_0)^2,
\end{displaymath}
for the $\beta$ model.  In these two specific cases the slope of the
exponential decay is given by $|\ln{\gamma_2}|$: this value is
preserved for almost all the energy distribution, until values of
${\mu}_j$ greater than $\mu_2$ appear in the tail.  These higher order
anomalies are the ones producing resonant tail peaks: the factor
$\mu_n$ in (\ref{3.firstest}) amplifies those modes whose frequencies
are almost resonant with $\omega_{q_0}$.  This effect is further
influenced by $\omega_{nq_0}$, which, depending on its value, can
weaken or strengthen the amplitudes (see section 3.4 and
Fig.~\ref{fig3}).  An unavoidable limitation is that the sequence
$\mu_j$ is constant with respect to the perturbation parameters
$\alpha$ and $\mathcal{E}_{q_0}$. Thus it does not take a possible
change of position and intensity of the peaks in the energy
distribution into account.

In the following, we will recall the basic ingredients of the
Poincar\'e-Lindstedt method. Then we present the rules used to predict
the cascade of modes excited in any $q_0$-breather. Proposition
\ref{3.prop1} can be suitably extended to any homogeneous potential.
It is strongly related to some symmetries of the system, as explained
in Appendix A and B (see also \cite{RINKinv}).  Finally we move to the
main general statement about localization, performing the proof for
$\alpha,\,\beta$ and a combined $\alpha-\beta$ model.  We finish with
an application to the high-frequency-mode case $|q_0-N| \ll N$.

%-----------------------------------------------------------------------------
%POINCARE'-LINDSTEDT

\subsection{The Poincar\'e-Lindstedt method.}

The Poincar\'e-Lindstedt expansion is a classical perturbation method
used to continue a periodic orbit with respect to a small perturbation
parameter, when fixing the amplitude (or the energy) of the
system. The crucial idea is to expand both the solution and the
modified frequency with respect to this parameter: this is essential
to kill secular terms which appear in the recursive scheme.

For example, consider directly the $\alpha$-FPU system: we start
expanding in $\sigma=\frac{\alpha}{\sqrt{2N+2}}$ both the normal mode
coordinates and the new orbit's frequency
\begin{displaymath}
Q_q(t) = \sum_{j\geq 0}{\sigma^j Q_q^{(j)}(t)},\qquad\qquad
\Omega(\sigma) = \sum_{j\geq 0}{\sigma^j\omega^{(j)}}.
\end{displaymath}
We define a new stretched time variable $\tau = \Omega(\sigma)t$ and we 
rewrite the equation of motion with respect to the new time derivative $\frac{d}{d\tau}$
\begin{equation}
\label{3.eqTau}
{\Omega(\sigma)^2\frac{d^2}{d\tau^2}Q_q}(\tau)+\omega_q^2 Q_q(\tau) = -\sigma\sum_{1\leq l,
  m\leq N}{\omega_q\omega_l\omega_m Q_l(\tau) Q_m(\tau) B_{qlm}}.
\end{equation}
If we expand (\ref{3.eqTau}) with respect to $\sigma$ we get an
infinite sequence of equations for every term $Q_q$
\begin{eqnarray}
\label{3.eqTau2}
\sum_{h=0}^n{\Omega_h^2 \frac{d^2}{d\tau^2}Q_q^{(n-h)}} &=& -
\omega_q^2 Q_q^{(n)} - \cr
&-& \omega_q\sum_{l,m=1}^N{\omega_l\omega_m B_{qlm}\sum_{h=0}^{n-1}{Q_l^{(h)}Q_m^{(n-1-h)}}},
\end{eqnarray}
where
\begin{displaymath}
  \Omega^2(\sigma)=\sum_{h\geq 0}{\sigma^h\Omega_h^2},\qquad\qquad \Omega_h^2 =
\sum_{i=0}^h{\omega^{(i)}\omega^{(h-i)}}.
\end{displaymath}
We need solutions of (\ref{3.eqTau2}) which are $2\pi$-periodic in $\tau$, so
bounded. The initial condition will be chosen so to guarantee this
requirement, and the term $\omega^{(j)}$ will be used to prevent the
existence of secular terms (unbounded motions) which could appear in the
equation of the continued mode.

%------------------------------------------------------------------------------
%ANY Q_0

\subsection{Cascade of modes.}

The QBs computed in Fig.~\ref{fig3} present an interesting feature
that was already visible in \cite{FlaIvaKan05}:  only a small
subset of modes is taking part to the dynamics, the other being essentially
at rest. So, for a full description of such solutions, we aim at understanding
which mechanism is responsible for this modes' selection.
This is also essential in order to get some straightforward result from
equation (\ref{3.eqTau2}). The main rule is that, 
at any $n$-step of perturbation theory, just one new mode $\overline{q}_n$ is excited: 
this corresponds to the only mode in the interval $[1,N]$ whose frequency equals
$2|\sin{(\frac{nq_0\pi}{2N+2})}|$. The count of $\overline{q}_n$ stops when either $nq_0$
is a multiple of $N+1$ or when $nq_0$ and $(n-1)q_0$ have the same
frequency (i.e. they are symmetric with respect to one of the two dispersion edges). 
A more rigorous claim is the following (for a sketch of the proof see Appendix A):
\begin{proposition}
\label{3.prop1}
At any $k-1$ perturbation order, only one new mode $\overline{q}_k$
with frequency $2|\sin{(\frac{kq_0\pi}{2N+2})}|$ is excited. This mode
reads
\begin{equation}
\overline{q}_k = 
\begin{cases}
{
(kq_0)\,{\rm mod}(2N+2),\qquad\qquad\qquad (kq_0)\,{\rm mod}(2N+2)<N+1\cr
2N+2-(kq_0)\,{\rm mod}(2N+2),\qquad {\rm otherwise}\cr
}
\end{cases}
\end{equation}
Thus it holds
\begin{displaymath}
Q_{\overline{q}_k}^{(m)}(t)=0\qquad\forall m<k-1.
\end{displaymath}
\end{proposition}

The above proposition exactly predicts which modes are actually
excited by the QB's dynamics. These orbits are strictly related to
some of those invariant submanifolds whose existence has been proved
in \cite{RINKinv} using discrete symmetries of the system. However,
differently from those arguments, the above proposition provides also
an order according to which the modes take part in the evolution.

The final scenario is the following. Take $1\leq q_0\leq N$ and
$g_0:=gcd(q_0,2N+2)$ (as in Proposition \ref{3.prop2}).  For $g_0=1$
the submanifold and the QB are simply the whole mode space.  For
$g_0\geq 2$ and $\frac{2N+2}{g_0}\not\in\mathbb{N}$ the submanifold
has lower dimension that the number of modes, at the same time it can
not be embedded in a subsystem with fixed boundary conditions. Hence,
the QB does not involve all the modes, but is not a rescaled solution
of a lower dimensional system.  For $g_0\geq 2$ and $g_0=gcd(N+1,q_0)$
the submanifold is embedded in a fixed boundary condition subsystem
with $\frac{N+1}{g_0}$ particles and the QB is the
$\frac{q_0}{g_0}$-mode rescaled solution, constructed as in
\cite{KanFlaIvaMsh06}.

%------------------------------------------------------------------------------
%EXPONENTIAL LOCALIZATION

\subsection{Exponential localization}
Next we prove the exponential decay of the harmonic energies for a
generic QB. This proof makes use of the sequence of small denominators
previously denoted as $\eta_{n,m}$, since it naturally arises in the
PL method. The Fourier coefficients in the PL scheme, at any new step,
are naturally defined via a convolution product.  Since $l^2_b$ is not
closed with respect to this product, the choice of a $\luno{\cdot}\geq
\|\cdot\|_{l^2_b}$ norm seems natural to control the main physical
quantities.  It is then useful to recall the following

\begin{lemma}
\label{3.lemmaconv}
Let $l^1_b$ denote the Banach space of absolutely convergent
bi-infinite sequences, endowed with the norm
$\luno{a}:=\sum_{j\in\mathbb{Z}}{|a_j|}$. Then $l^1_b$ is a Banach
algebra with respect to the convolution product
\begin{displaymath}
( a \star b)_m = \sum_{i+j=m}{a_i b_j}
\end{displaymath}
and 
\begin{equation}
\label{3.luno}
\luno{ a \star b} \leq \luno{ a}\luno{ b} 
\end{equation}
holds.
\end{lemma}

\smunp

\subsubsection{The $\alpha$ model.}
We develop an estimate for the averaged harmonic energies of the
leading terms $Q_{{\overline q}_k}^{(k-1)}$, based on the scheme
presented at the beginning of this section. By applying
Prop. \ref{3.prop1} and with the same notation for the modes, equation
(\ref{3.eqTau2}) reads
\begin{eqnarray}
\label{3.starteq}
\omega_{q_0}^2 \frac{d^2}{d\tau^2}Q_{{\overline q}_k}^{(k-1)} &=&
-\omega_{kq_0}^2 Q_{{\overline q}_k}^{(k-1)} -\cr
&\pm&\omega_{kq_0}\sum_{l=1}^{k-1}{\omega_{lq_0}\omega_{(k-l)q_0}
Q_{{\overline q}_l}^{(l-1)}Q_{{\overline
q}_{k-l}}^{(k-1-l)}}.
\end{eqnarray}
At the first step ($k=2$) 
\begin{displaymath}
Q_{{\overline q}_2}^{(1)}(\tau) = \pm
\frac{\omega_{2q_0}\omega_{q_0}^2 A_{q_0}^2}{2}
\left[\frac1{\omega_{2q_0}^2} +
\frac{\cos(2\tau)}{\omega_{2q_0}^2-4\omega_{q_0}^2}\right].
\end{displaymath}
We define the quantities
\begin{displaymath}
\overline\eta_{2,l} := |\omega_{2q_0}^2-l^2\omega_{q_0}^2|,\qquad
{\eta}_2:=\min_{l=0,\ldots,2}{\{\eta_{2,l}\}},\qquad
\mu_2:=\frac1{{\eta}_2}.
\end{displaymath}
The solution becomes 
\begin{displaymath}
Q_{{\overline q}_2}^{(1)}(\tau) = \sum_{m=-2}^2{\hat
C_{2,m}\exp{(im\tau)}},\qquad \hat C_{2,m} = \hat C_{2,-m},\qquad C_{2,1} = 0,
\end{displaymath}
and the Fourier sequence satisfies
\begin{displaymath}
\luno{\hat C_2}\leq\mu_2\luno{\hat C_1}^2 c_2,\qquad\qquad c_2=\omega_{2q_0}\omega_{q_0}^2.
\end{displaymath}
Now, we state and prove the main

%TEOREMA
\begin{theorem}
\label{3.alfateo}
Define for any $k\geq 2$ the following quantities:
\begin{eqnarray*}
\overline\eta_{k,l} &:=& |\omega_{kq_0}^2-l^2\omega_{q_0}^2|,\qquad\qquad
{\eta}_k := \min_{l=k,k-2,,\ldots}{\{\overline\eta_{k,l}\}},\\ \mu_k &:=&
\frac{1}{{\eta}_k},\qquad\qquad\qquad\qquad \overline{\mu}_k=\max_{l=2,\ldots,k}{\{\mu_l\}}.
\end{eqnarray*}
Then, the leading solution $Q_{kq_0}^{(k-1)}(\tau)$ reads
\begin{displaymath}
Q_{{\overline q}_k}^{(k-1)}(t) = \sum_{m=-k}^k{\hat C_{k,m}\exp{(i
m\Omega t)}},\qquad\hat C_{k,m}=0, \, m=k-1,\ldots 1-k,
\end{displaymath}
and there exists a positive constant $a>1$ such that the Fourier
sequence $\hat C_k$ fulfills the following estimate
\begin{displaymath}
\luno{\hat C_k}\leq {\mu_k}\left(\overline\mu_{k-1}\right)^{k-2}
\luno{\hat C_1}^k c_k,\qquad\qquad c_k<\omega_{q_0}^ka^{k-1}.
\end{displaymath}
Moreover, if $\mathcal{E}_{q_0} = \frac{E_{q_0}}{N+1}$ represents the
energy density of the mode $q_0$, there exists a threshold
$\mathcal{E}^*_{q_0}$ such that, for $\mathcal{E}_{q_0} <
\mathcal{E}^*_{q_0}$, the leading averaged harmonic energy fulfills
\begin{displaymath}
\langle E_{\overline q_k}\rangle < E_{q_0}(\omega_{nq_0}^2+n^2\omega_{q_0}^2)
\left(\frac{\mu_k}{\overline\mu_{k-1}} \right)^2
e^{-(1-k)|\ln{\gamma_k}|},
\end{displaymath}
with
\begin{displaymath}
\gamma_k = 2\left(\alpha a {\overline\mu_{k-1}}\right)^2
\mathcal{E}_{q_0}.
\end{displaymath}
\end{theorem}

%-------------------
%proof
\begin{proof}
By induction. The case $k=2$ has been already proved. Notice that it
is not necessary to define $\overline\mu_1$, since its exponent is
$0$. Suppose the statement is true for $k=n-1$ and let us prove it for
$k=n$. The time-dependent force is composed of harmonics with
frequencies running from $0$ to $n$
\begin{displaymath}
Q_{{\overline q}_l}^{(l-1)}(t) Q_{{\overline q}_{n-l}}^{(n-1-l)}(t) =
\sum_{m=-n}^n{(\hat C_l \star\hat C_{n-l})_m\exp{(i m\Omega t)}},
\end{displaymath}
with
\begin{displaymath}
(\hat C_l \star\hat C_{n-l})_m = (\hat C_l \star\hat
  C_{n-l})_{-m},
\end{displaymath}
and
\begin{displaymath}
(\hat C_l \star\hat C_{n-l})_m=0,\qquad\qquad m=n-1,\ldots,1-n. 
\end{displaymath}
From the definition of the force Fourier coefficient
\begin{displaymath}
\hat F^{(n-1)}_{n,m} =
\sum_{l=1}^{n-1}{B_{\overline{q}_n,\overline{q}_l,\overline{q}_{n-l}}(\hat
C_l \star\hat C_{n-l})_m \omega_{lq_0}\omega_{(k-l)q_0}}
\end{displaymath}
we can expand the time dependent force
\begin{displaymath}
F^{(n-1)}_{\overline q_n}(t) = -\omega_{nq_0}\sum_{m=-n}^n{\hat
  F^{(n-1)}_{n,m}\exp{(i m\Omega t)}},
\end{displaymath}
and write the solution as
\begin{eqnarray*}
Q^{(n-1)}_{\overline q_n}(t) &=& -\omega_{nq_0}\sum_{m=-n}^n{\frac{\hat
    F^{(n-1)}_{n,m}}{\omega_{nq_0}^2-m^2\omega_{q_0}^2}\exp{(i m\Omega t)}} =\cr
    &=&\sum_{m=-n}^n{\hat C_{n,m}\exp{(i m\Omega t)}}.
\end{eqnarray*}
Now we move to the estimate:
\begin{eqnarray*}
\luno{\hat C_n} &\leq & \frac{\omega_{nq_0}}{
    \eta_n}\sum_{m=-n}^n{\frac{\eta_n}{\overline\eta_{n,m}}|\hat
    F^{(n-1)}_{n,m}|} \leq \\ &\leq& \frac{\omega_{nq_0}}{
    \eta_n} \sum_{l=1}^{n-1}{\omega_{lq_0}\omega_{(n-l)q_0}\luno{\hat
    C_l \star\hat C_{n-l}}} \leq \\ &\leq& \frac{\omega_{nq_0}}{
    \eta_n} \sum_{l=1}^{n-1}{\omega_{lq_0}\omega_{(n-l)q_0}\luno{\hat
    C_l}\luno{\hat C_{n-l}}}<\\  &<&\frac{\omega_{nq_0}}{\eta_n}\luno{\hat C_1}^n
    \left(\overline\mu_{n-1}\right)^{n-2}
    \sum_{l=1}^{n-1}{\omega_{lq_0}\omega_{(n-l)q_0}c_l c_{n-l}} = \\
    &=& \mu_n\luno{\hat C_1}^n
    \left(\overline\mu_{n-1}\right)^{n-2}c_n,
\end{eqnarray*}
where we have defined by recurrence
\begin{equation}
c_n := \omega_{nq_0}\sum_{l=1}^{n-1}{\omega_{lq_0}\omega_{(n-l)q_0}c_l c_{n-l}}.
\end{equation}
The values of the first elements $c_2,c_3,c_4,\ldots$ 
suggest the possibility to collect a factor $\omega_{q_0}^n$ from any term $c_n$: 
indeed, if we set $d_2=\omega_{2q_0}$ and
\begin{displaymath}
  d_n
  :=\omega_{nq_0}\left(2d_{n-1}\omega_{(n-1)q_0}+\sum_{j=2}^{n-2}{\omega_{lq_0}\omega_{(n-l)q_0}d_l
  d_{n-l}}\right),
\end{displaymath}
then $c_n=\omega_{q_0}^n d_n$, with $d_n<a^{n-1}$.  The bound on the
amplitudes $c_n<\omega_{q_0}^{n}a^{n-1}$, which is proved in the
Appendix C, allows to estimate the averaged leading terms of the
harmonic energies (\ref{3.avharm}) 
\begin{displaymath}
\langle E_{\overline q_n}\rangle =
\frac{\sigma^{2n-2}}{\tau}\int_0^\tau{\frac12 \left[
\omega_{nq_0}^2(Q_{{\overline q}_n}^{(n-1)})^2(t) + (P_{{\overline
q}_n}^{(n-1)})^2(t)\right]dt},
\end{displaymath}
where 
\begin{displaymath}
P_{{\overline q}_n}^{(n-1)}(t) = \sum_{m=-n}^n{i m
\omega_{q_0}\hat C_{n,m}\exp{(i m\Omega t)}}
\end{displaymath}
is obtained by differentiating $Q_{{\overline q}_n}^{(n-1)}(t)$ with
respect to time $t$ and recalling that $\Omega = \omega_{q_0} +
\mathcal{O}(\sigma)$. The period-average yields to
\begin{eqnarray*}
\langle E_{\overline q_n}\rangle &=& \sigma^{2n-2}\sum_{|m|\leq
n}{\frac12(\omega_{nq_0}^2+m^2\omega_{q_0}^2)|\hat C_{n,m}|^2} <\\ &<&
\sigma^{2n-2}\frac12(\omega_{nq_0}^2+n^2\omega_{q_0}^2) \sum_{|m|\leq
n}{|\hat C_{n,m}|^2}\leq\\ &\leq&
\sigma^{2n-2}\frac12(\omega_{nq_0}^2+n^2\omega_{q_0}^2)\luno{\hat
C_n}^2.
\end{eqnarray*}
By means of the previous estimate on $\luno{\hat C_n}$ we get
\begin{displaymath}
E_{\overline q_n}^{(n-1)} \leq
E_{q_0}(\omega_{nq_0}^2+n^2\omega_{q_0}^2)
\left(\frac{\mu_n}{\overline\mu_{n-1}} \right)^2
e^{-(1-n)|\ln{\gamma_n}|},
\end{displaymath}
which is an exponential decay until
\begin{equation}
\label{3.gammadef}
\gamma_n = 2\left(\alpha a \overline\mu_{n-1}\right)^2 \mathcal{E}_{q_0} <
1,\qquad\qquad\forall n=1,\ldots,N-1.
\end{equation}
If we define $\mu=\max{\overline\mu_n}$, the condition
\begin{equation}
\gamma < 1 \qquad \Leftrightarrow \qquad \mathcal{E}_{q_0} <
\mathcal{E}_{q_0}^*:={2\left({a \alpha \mu}\right)^{-2}}
\end{equation}
repesents the $q_0$ specific energy's threshold for the exponential
decay.

\begin{flushright}
$\blacksquare$
\end{flushright}
\end{proof}

%------------------------------\beta model----------------------------

\subsubsection{The $\beta$ model.}

For the $\beta$-model we give the main ideas how
to extend the perturbation estimates performed for the $\alpha$ model. The
quartic part of the potential can be rewritten as
\begin{displaymath}
{\frac{\beta}{4}}\sum_{j=0}^N{(x_{j+1}-x_j)^4}={\frac{\beta}{8(N+1)}}\sum_{1\leq
  i, j, h, k\leq N}{\omega_i\omega_j\omega_h\omega_k
  Q_i Q_j Q_h Q_k B_{ijhk}},
\end{displaymath}
where we set $ B_{ijhk} = \sum_{\sigma_{1,2,3}=\pm 1}{\Delta_{i
+\sigma_1 j +\sigma_2 h +\sigma_3 k}}$, with
\begin{equation}
\label{4.qrtrules}
\Delta_{i\pm j \pm h\pm k} = \left\{\matrix{1, & i\pm j \pm h \pm k=0\cr -1, &
    i\pm j \pm h \pm k= \pm 2(N+1)\cr 0, & {\rm otherwise}}\right..
\end{equation}
The equation of motion for a normal mode then reads
\begin{equation}
\label{4.motion}
\ddot{Q_q}+\omega_q^2 Q_q = -\frac{\beta}{2(N+1)}\sum_{1\leq i, j, h\leq N}{\omega_q\omega_i\omega_j\omega_h
  Q_i Q_j Q_h B_{qijh}}.
\end{equation}
The small parameter used in the Taylor expansion is 
$\sigma = \frac{\beta}{2N+2}$. With this choice, the equation at
every order $s$ becomes
\begin{eqnarray*}
\sum_{r=0}^s{\Omega_r^2 \frac{d^2}{d\tau^2}Q_q^{(s-r)}} &=& - \omega_q^2 Q_q^{(s)} -\cr
&-&\omega_q\sum_{i,j,h=1}^N{\omega_i\omega_j\omega_h
  B_{qijh}\sum_{l+m+n=s-1}{Q_i^{(l)} Q_j^{(m)} Q_h^{(n)}}}.
\end{eqnarray*}
Following both \cite{FlaIvaKan05} and Proposition \ref{3.prop1} we can prove

\begin{proposition}
\label{4.prop1}
At any $s-1$ perturbation order a new mode $\overline{q}_s$ with
frequency $2|\sin{\left(\frac{(2s-1)q_0\pi}{2N+2}\right)}|$ is
excited:
\begin{equation}
\overline{q}_s = 
\begin{cases}
{
(2s-1)q_0\,{\rm mod}(2N+2),\quad (2s-1)q_0\,{\rm mod}(2N+2)<N+1\cr
2N+2-(2s-1)q_0\,{\rm mod}(2N+2),\quad {\rm otherwise}\cr
}
\end{cases}
\end{equation}
Thus
\begin{displaymath}
Q_{\overline{q}_s}^{(m)}(t)=0\qquad\forall m<s-1.
\end{displaymath}
\end{proposition}

Consider $s=2$: the solution of  
\begin{displaymath}
\omega_{q_0}^2 \frac{d^2}{d\tau^2}Q_{\overline{q}_2}^{(1)} + \omega_{3q_0}^2
Q_{\overline{q}_2}^{(1)} =
\pm\omega_{3q_0} \omega_{q_0}^3 \left(Q_{q_0}^{(0)}\right)^3 
\end{displaymath}
is given by
\begin{displaymath}
Q_{\overline{q}_2}^{(1)}(\tau) = \pm \omega_{3q_0}\omega_{q_0}^3 A_{q_0}^3 
\left[\frac34\frac{\cos(\tau)}{\omega_{3q_0}^2 - \omega_{q_0}^2} + \frac14\frac{\cos(3\tau)}{\omega_{3q_0}^2-9\omega_{q_0}^2}\right],
\end{displaymath}
or, after expanding cosine terms
\begin{displaymath}
Q_{\overline{q}_2}^{(1)}(\tau) = \sum_{m=-3}^3{\hat
  C_{2,m}\exp{(im\Omega t)}},\qquad\hat C_{2,2m}=0,\qquad\hat
  C_{2,m}=\hat C_{2,-m}.
\end{displaymath}
We define the following quantities
\begin{displaymath}
\eta_{2,l} := |\omega_{3q_0}^2-l^2\omega_{q_0}^2|,\qquad
\overline{\eta_2}:=\min_{l=1,3}{\{\eta_{2,l}\}},\qquad
\mu_2:=\frac{1}{\overline{\eta_2}},
\end{displaymath}
then 
\begin{displaymath}
\luno{\hat C_2}\leq  \mu_2 \luno{\hat C_1}^3 c_2,\qquad\qquad c_2=\omega_{3q_0}\omega_{q_0}^3.
\end{displaymath}

%TEOREMA
\begin{theorem}
\label{3.betateo}
Define for any $s\geq 2$ the following quantities:
\begin{eqnarray*}
\overline\eta_{s,l} &:=& |\omega_{(2s-1)q_0}^2-l^2\omega_{q_0}^2|,\qquad
{\eta_s} := \min_{l=1,3,\ldots,2s-1}{\{\eta_{s,l}\}},\\ \mu_s &:=&
\frac1{{\eta_s}},\qquad\qquad\qquad\qquad \overline{\mu}_s=\max_{l=2,\ldots,s}{\{\mu_s\}}.
\end{eqnarray*}
Then, the leading solution $Q_{\overline q_s}^{(s-1)}(t)$ reads
\begin{equation}
Q_{\overline q_s}^{(s-1)}(t) =
\sum_{m=-2s+1}^{2s-1}{\hat C_{s,m}\exp{(i m\Omega t)}},\qquad\qquad \hat
C_{s,2m} = 0, 
\end{equation}
and there exists a positive constant $a>1$ such that the Fourier
sequence $\hat C_s$ fulfills the following estimate
\begin{displaymath}
\luno{\hat C_s}\leq \mu_s\left(\overline\mu_{s-1}\right)^{s-2} \luno{\hat
C_1}^{2s-1} c_s,\qquad\qquad c_s<\omega_{q_0}^{2s-1} a^{s-1}.
\end{displaymath}
Moreover, if $\mathcal{E}_{q_0} = \frac{E_{q_0}}{N+1}$ represents the
energy density of the mode $q_0$, there exists a threshold
$\mathcal{E}^*_{q_0}$ such that, for $\mathcal{E}_{q_0} <
\mathcal{E}^*_{q_0}$
\begin{displaymath}
\langle E_{\overline q_s}\rangle <
E_{q_0}(\omega_{(2s-1)q_0}^2+(2s-1)^2\omega_{q_0}^2)
\left(\frac{\mu_s}{\overline\mu_{s-1}} \right)^2
e^{-2(1-s)|\ln{\gamma_s}|},
\end{displaymath}
with
\begin{displaymath}
\gamma_s = \beta a {\overline\mu_{s-1}}\mathcal{E}_{q_0} .
\end{displaymath}
\end{theorem}

\begin{proof}

We give some hints how to modify the proof of Theorem
\ref{3.alfateo}. The nonlinearity in the equation of motion for
$Q_{\overline q_s}^{(s-1)}$ is
\begin{displaymath}
\omega_{(2s-1)q_0}\sum_{l+m+n=s+1}{\omega_{(2l-1)q_0}\omega_{(2m-1)q_0}\omega_{(2n-1)q_0}
Q_{\overline q_l}^{(l-1)} Q_{\overline q_m}^{(m-1)} Q_{\overline q_n}^{(n-1)}},
\end{displaymath}
where $l,m,n\geq 1$.  We can define the force Fourier coefficient
\begin{displaymath}
\hat F^{(s-1)}_{s,h} =
  \sum_{l+m+n=s+1}{B_{\overline{q}_s,\overline{q}_l,\overline{q}_m
  \overline{q}_n}(\hat C_l \star\hat C_m \star\hat C_n})_h
  \omega_{(2l-1)q_0}\omega_{(2m-1)q_0}\omega_{(2n-1)q_0}
\end{displaymath}
 and obtain the time dependent force
\begin{displaymath}
F^{(s-1)}_{\overline q_s}(t) =
-\omega_{(2s-1)q_0}\sum_{h=-2s+1}^{2s-1}{\hat F^{(s-1)}_{s,h}\exp{(i h\Omega t)}}.
\end{displaymath}
This, as before, yields the solution
\begin{eqnarray*}
Q^{(s-1)}_{\overline q_s}(t) &=&
-\omega_{(2s-1)q_0}\sum_{h=-2s+1}^{2s-1}{\frac{\hat
      F^{(s-1)}_{s,h}}{\omega_{(2s-1)q_0}^2-h^2\omega_{q_0}^2}\exp{(i h\Omega t)}} =\cr
  &=&\sum_{h=-2s+1}^{2s-1}{\hat C_{s,h}\exp{(i h\Omega t)}}.
\end{eqnarray*}
The conclusion of the proof follows Theorem \ref{3.alfateo}.
\begin{flushright}
$\blacksquare$
\end{flushright}
\end{proof}

%---------------------------------------------------------------------------
%alpha-beta model

\subsubsection{The $\alpha-\beta$ model with $\beta=\frac23\alpha^2$.}

We finally consider the $\alpha-\beta$ model, under the restriction
$\beta=\frac23\alpha^2$. This dependence is due to the need of
expanding the solution with respect to one perturbative parameter
only. Moreover, with this choice the potential coincides with the
fourth truncation order of the Toda potential and the scheme performed
can be adapted to any order.

In this case, defining $\sigma=\frac{\alpha}{\sqrt{2N+2}}$, the
equation at any order $s-1$ reads

\begin{eqnarray*}
\sum_{r=0}^{s-1}{\Omega_r^2 \frac{d^2}{d\tau^2}Q_q^{(s-1-r)}} &=& -
\omega_q^2 Q_q^{(s-1)} -\omega_q\sum_{i,j=1}^N{\omega_i\omega_j
B_{qij}\sum_{l+m=s-2}{Q_i^{(l)} Q_j^{(m)} }}-\\
&-&\frac23\omega_q\sum_{i,j,h=1}^N{\omega_i\omega_j\omega_h
C_{qijh}\sum_{l+m+n=s-3}{Q_i^{(l)} Q_j^{(m)} Q_h^{(n)}}}.
\end{eqnarray*}

As in the $\alpha$ model, the set of modes involved in
the dynamics is selected according to Proposition \ref{3.prop1} (see
again Appendix A)
\begin{proposition}
\label{4.prop2}
At any $k-1$ perturbation order, only one new mode $\overline{q}_k$
with frequency $2|\sin{(\frac{kq_0\pi}{2N+2})}|$ is excited:
\begin{equation}
\overline{q}_k = 
\begin{cases}
{
(kq_0)\,{\rm mod}(2N+2),\qquad\qquad\qquad (kq_0)\,{\rm mod}(2N+2)<N+1\cr
2N+2-(kq_0)\,{\rm mod}(2N+2),\qquad {\rm otherwise}\cr
}
\end{cases}
\end{equation}
Thus 
\begin{displaymath}
Q_{\overline{q}_k}^{(m)}(t)=0\qquad\forall m<k-1.
\end{displaymath}
\end{proposition}

The exponential decay is similar to the one obtained for the 
$\alpha$ model, except for the coefficient $c_s$ which is responsible for the
accumulation of frequencies:

\begin{theorem}
\label{3.alfabetateo}
Define for any $s\geq 2$ the following quantities:
\begin{eqnarray*}
\overline\eta_{s,l} &:=& |\omega_{sq_0}^2-l^2\Omega_{q_0}^2|,\qquad\qquad
{\eta}_s := \min_{l=s,s-2,,\ldots}{\{\eta_{s,l}\}},\\ \mu_s
&:=&
\frac1{{\eta}_s},\qquad\qquad\qquad\qquad
\overline{\mu}_s=\max_{l=2,\ldots,s}{\{\mu_l\}}.
\end{eqnarray*}
Then any solution $Q_{{\overline q}_s}^{(s-1)}(t)$ reads
\begin{displaymath}
Q_{{\overline q}_s}^{(s-1)}(t) = \sum_{m=-s}^s{\hat C_{s,m}\exp{(i
    m\Omega t)}},\qquad\hat C_{s,m} =
    0,\,m=s-1,\ldots,1-s
\end{displaymath}
and there exists a positive constant $a>1$ such that the Fourier
sequence $\hat C_s$ fulfills the following estimate
\begin{displaymath}
\luno{\hat C_s}\leq \mu_s\left(\overline\mu_{s-1}\right)^{s-2} \luno{\hat
C_1}^s c_s,\qquad\qquad c_s< \omega_{q_0}^s a^{s-1}.
\end{displaymath}
Moreover, if $\mathcal{E}_{q_0} = \frac{E_{q_0}}{N+1}$ represents the
energy density of the mode $q_0$, there exists a threshold
$\mathcal{E}^*_{q_0}$ such that, for $\mathcal{E}_{q_0} <
\mathcal{E}^*_{q_0}$, holds true
\begin{displaymath}
\langle E_{\overline q_s}\rangle <
E_{q_0}(\omega_{sq_0}^2+s^2\omega_{q_0}^2)
\left(\frac{\mu_s}{\overline\mu_{s-1}} \right)^2
e^{-(1-s)|\ln{\gamma_s}|},
\end{displaymath}
with
\begin{displaymath}
\gamma_s = 2\left(\alpha a {\overline\mu_{s-1}}\right)^2
\mathcal{E}_{q_0}.
\end{displaymath}
\end{theorem}

\begin{proof}
By induction we consider the step $\overline q_s$. The quadratic and
cubic force are composed of terms like
\begin{eqnarray*}
Q_{{\overline q}_l}^{(l-1)} Q_{{\overline q}_{m}}^{(m-1)} &=&
\sum_{r=-s}^s{(\hat C_l \star\hat C_{m})_r\exp{(i r\Omega
    t)}},\\
Q_{{\overline q}_l}^{(l-1)} Q_{{\overline q}_{m}}^{(m-1)}Q_{{\overline q}_{n}}^{(n-1)} &=&
\sum_{r=-s}^s{(\hat C_l \star\hat C_{m}\star\hat C_n)_r\exp{(i r\Omega
    t)}},
\end{eqnarray*}
with respectively $l+m=s$ and $l+m+n=s$; observe that 
\begin{displaymath}
(\hat C_l \star\hat C_{m})_r=(\hat C_l \star\hat C_{m}\star\hat
  C_n)_r=0,\qquad\qquad r=s-1,s-3,\ldots,1-s.
\end{displaymath}
Since in both the cases the harmonics run from $-s$ to $s$, it is
possible to define the force Fourier coefficient as
\begin{eqnarray*}
\hat F^{(s-1)}_{s,r} &=& \sum_{l=1}^{s-1}{(\hat C_l \star\hat
  C_{s-l})_r \omega_{lq_0}\omega_{(s-l)q_0}}+\\ &+&
  \sum_{l=1}^{s-2}{(\hat C_l \star \hat C_m \star \hat C_{s-l-m})_r
  \omega_{lq_0}\omega_{mq_0}\omega_{(s-l-m)q_0}}\\
\end{eqnarray*}
and expand the time dependent force
\begin{displaymath}
F^{(s-1)}_{\overline q_s} = -\omega_{sq_0}\sum_{r=-s}^s{\hat F^{(s-1)}_{s,r}\exp{(i r\Omega t)}},
\end{displaymath}
which leads to the solution
\begin{displaymath}
Q^{(s-1)}_{\overline q_s} = -\omega_{sq_0}\sum_{r=-s}^s{\frac{\hat
    F^{(s-1)}_{s,r}}{\omega_{sq_0}^2-r^2\Omega_{q_0}^2}\exp{(i r\Omega t)}} =
    \sum_{r=-s}^s{\hat C_{s,r}\exp{(i r\Omega t)}}.
\end{displaymath}
It is remarkable that the both force contributions generate the same
small denominators.  The proof can be then easily finished, using a
new definition for $c_s$
\begin{displaymath}
c_s :=
\omega_{sq_0}\left(\sum_{l=1}^{s-1}{\omega_{lq_0}\omega_{(s-l)q_0}c_l
  c_{s-l}}+\sum_{l=1}^{s-2}{\omega_{lq_0}\omega_{mq_0}\omega_{(s-l-m)q_0}c_l
  c_m c_{s-l-m}} \right).
\end{displaymath}

\begin{flushright}
$\blacksquare$
\end{flushright}
\end{proof}

%---------------------application: high-mode-case--------------------------------

\subsubsection{Application: the high mode case.}

\begin{figure}
\includegraphics[width=0.9\columnwidth]{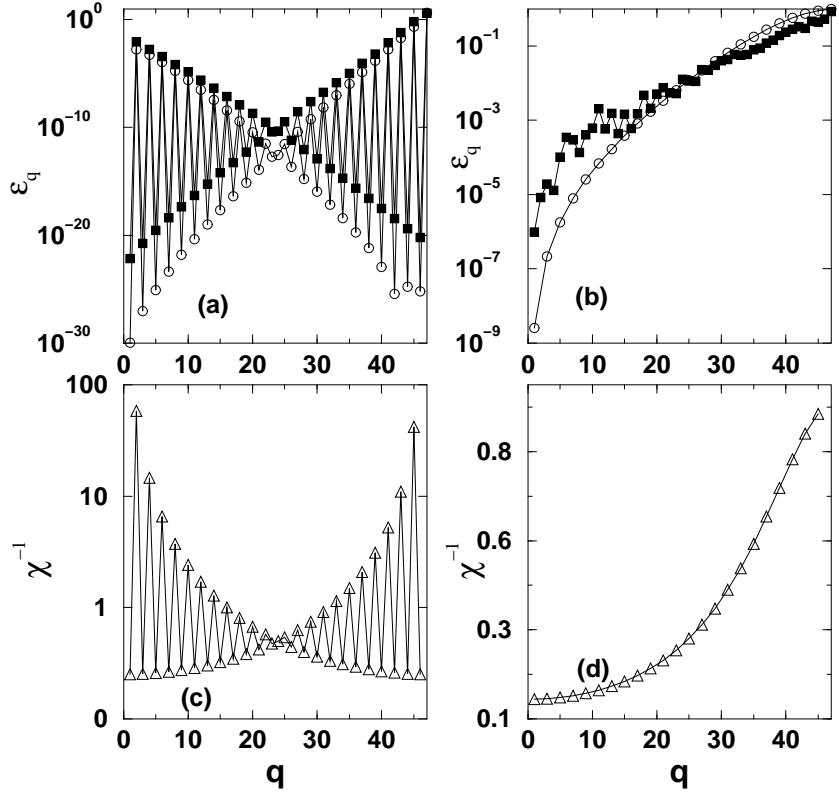}
\caption{Parameters: $N=47,\,\mathcal{E}=0.1$. (a): Energy of normal
modes in a QB (open symbols) and a natural packet (filled symbols,
waiting time $10^5$) for $q_0=47$ and $\alpha=0.25$. (b): Similar to
(a), but for $\beta=0.25$. (c): sequence $\chi_j^{-1}$ which
corresponds to case (a).  (d): sequence $\chi_j^{-1}$ which
corresponds to (b).}
\label{fig4} 
\end{figure}

We consider the case when $q_0$ corresponds to a large frequency,
i.e. $N-q_0\ll 1$.  As for a low frequency initial excitation, also
this special case is related to the equipartition issue. We recall,
for example, that in \cite{UllLC00} the energy flow from initially
high-frequency-mode excitation to low modes was studied, restricting
to a $\beta$-FPU model. The claim was that, due to the same mechanism
of resonance overlapping described in \cite{DelLL95}, the
equipartition time $T_{eq}$ scales as a power law with the inverse of
the energy density, $T_{eq}\sim\mathcal{E}^{-2}$.  However, the
authors didn't exclude also the possibility that, below a small energy
density threshold, the dependence $T_{eq}(\mathcal{E})$ could be even
slower than any power law.  The existence of exponentially long
waiting times to equipartition, scaling with the inverse of the energy
density, and the existence of two different time scales for the energy
spreading among modes, have been later numerically confirmed in
\cite{PP}.

While the impact of QBs on these observations is essentially the same
as for the low frequency case, a new element in the structure of the
QB itself arises.  For energy densities small enough, it is still
evident an exponential decay of harmonic energies starting from the
initially excited mode. However, a remarkable difference with the
low-mode case exists and it is related to the effect of the $\alpha$
term in the potential: if in the pure $\beta$ model the energy is
mainly shared among high frequency modes, in the $\alpha$ case the
energy distribution involves, from the very beginning of the
evolution, also low frequency modes. We compute QBs with high
frequencies, and plot their energy distributions in Fig.~\ref{fig4}.
The panels (a) and (b) of Fig.~\ref{fig4} show two $q$-breathers with
the same largest $q_0$, for the $\alpha$ and $\beta$ model,
respectively.  We also plot the energy distribution of the
corresponding natural packets. We again recover that both
distributions are remarkably similar.  Note that for the $\beta$-model
the modes decay monotonously with decreasing mode number.  However for
the $\alpha$-model a strong excitation of low frequency modes is
observed as well.

The two lower panels (c) and (d) in Fig.~\ref{fig4} represent the
$\chi_j$ sequences corresponding to the two QB periodic orbits.  Since
in the forthcoming part we want to obtain, from the general definition
of $\mathcal{E}^*$ in Theorem \ref{3.alfateo}, an approximated but
explicit dependence of the threshold energy density on
$\kappa_0:=\frac{q_0}{N+1}$, we will actually consider the sequence
$\eta_j$ instead, since $\eta_j$ is still very close to $\chi_j$. We
deal with the two initial small divisors $\eta_2$ and $\eta_3$,
arising respectively from the first and second PL step.  In the first
step, the low mode $2N+2-2q_0$ is coupled to $q_0$, but the forcing
term $Q^{(0)}_{q_0}(t)$ includes only even harmonics, thus generating
in the solution a small denominator $\frac1{\omega_{2q_0}^2}$.  We
expand $\omega_{2q_0}$ around $\pi$ and obtain the approximation
\begin{displaymath}
\omega_{2q_0}^2=4\sin^2{(\pi(1-\kappa_0))} \approx 4\pi^2(1-\kappa_0)^2.
\end{displaymath}
In the second step, instead, the mode $3q_0-2N-2$ is excited. The odd
harmonics of the time periodic force generate the small denominator
$\frac1{\omega_{3q_0}^2-\omega_{q_0}^2}$. Since both $\omega_{3q_0}$
and $\omega_{q_0}$ are very close to $2$, their difference is expected
to be quadratic in $(1-\kappa_0)$. We expand $\omega_{3q_0}$ in a
neighbourhood of $\frac32\pi$
\begin{displaymath}
\omega_{3q_0}=2\sin{(\frac32\pi\kappa_0)} =
2\cos{(\frac32\pi-\frac32\pi\kappa_0)}=2-\frac94\pi^2(1-\kappa_0)^2+{\rm h.o.t.},
\end{displaymath}
and $\omega_{q_0}$ around $\frac12\pi$
\begin{displaymath}
\omega_{q_0}=2\sin{(\frac12\pi\kappa_0)} =
2\cos{(\frac12\pi-\frac12\pi\kappa_0)}=2-\frac14\pi^2(1-\kappa_0)^2+{\rm h.o.t.},
\end{displaymath}
thus deducing
\begin{equation}
|\omega_{3q_0}^2-\omega_{q_0}^2|\approx 8\pi^2(1-\kappa_0)^2.
\end{equation}
So, in agreement with panel (c) in Fig.~\ref{fig4}, we have finally
explained the relationship $\eta_2>\eta_3$. Thus we may expect
$\langle E_{\overline q_2}\rangle > \langle E_{\overline
q_3}\rangle$. To compare panel (c) with panel (a) in Fig.~\ref{fig4}
we recall that the averaged harmonic energies $\langle E_{\overline
q_s}\rangle$ are also corrected by the coefficients
$(\omega_{sq_0}^2+s^2\omega_{q_0}^2)$. Finally, from the previous
local approximation of strong nearby resonances, one can deduce
\begin{displaymath} 
\mathcal{E}^*_{q_0}\approx \alpha^{-2}\pi^4 (1-\kappa_0)^4.
\end{displaymath} 
It is straightforward to deal with the $\beta$ case and obtain
from Theorem \ref{3.betateo}
\begin{displaymath} 
\mathcal{E}^*_{q_0}\approx \beta^{-1}\pi^2 (1-\kappa_0)^2.
\end{displaymath} 

%---------------------------------------------------------------------------
%---------------------------------------------------------------------------
%
%                     CONCLUSIONS
%
%---------------------------------------------------------------------------
%---------------------------------------------------------------------------

\section{Conclusions.}
We have studied the energy localization of a generic Lyapunov orbit,
called $q$-breather, in the pure $\alpha$ and $\beta$ FPU model. We
have proved, thus generalizing a result from \cite{FlaIvaKan05}, that
any $q_0$-breather exhibits an exponential localization in the Fourier
space if the $q_0$-mode energy density $\mathcal{E}_{q_0}$ is below a
threshold $\mathcal{E}_{q_0}^*$ which depends only on the coupling
parameter (either $\alpha$ or $\beta$) and on the wave number
$\kappa_0=\frac{q_0}{N+1}$. The specific value of
$\mathcal{E}_{q_0}^*$ suggests that a $\beta$-QB will delocalize for
higher energies. Moreover, the exponential decay of amplitudes in a QB
runs through a subset of normal modes: we revealed the algebraic rule
which selects the modes involved in the QB dynamics, thus explaining
their order in the perturbation decay. This rule is strictly linked to
the discrete symmetries of the system. Any QB lies on one of those
invariant submanifolds described, for example, in \cite{MR1464250} and
\cite{RINKinv}.

The possibility to evaluate the Poincar\'e-Lindstedt (PL) perturbation
scheme at any order, allows a qualitative explanation of the
previously observed resonant peaks in the tail of the energy
distribution.  In the numerical experiments on natural packets with
initial low frequency modes, we notice anomalies in the tail of the
energy distribution. They pump energy from the natural packet core
into the tail. We think that this effect is of central relevance to
the pathway to equipartition. Such anomalies are also characteristic
of any QB, even if in the low fequency mode simulations (see
Fig.~\ref{fig1} and Fig.~\ref{fig3}) they are more visible. While
these peaks have been plotted (but not discussed) already in
\cite{FlaIvaKan05}, we have developed a scheme which is capable of
explaining the positions of these peaks.  We have shown that ther is a
direct connection between the position and intensity of these peaks
and nonlinear resonances involving the QB's frequency $\Omega_{q_0}$
\begin{equation}
\label{5.trueres}
\overline\chi_{n,m} = |\omega_{nq_0}^2 - m^2\Omega_{q_0}^2|.
\end{equation}
In the limit of very small perturbations, the sequence
(\ref{5.trueres}) reduces to
\begin{equation}
\label{5.res}
\overline\eta_{n,m} = |\omega_{nq_0}^2 - m^2\omega_{q_0}^2|,
\end{equation}
which is the one arising in the PL scheme. We have used the latter one
for a rigorous analytical approach. We still lack a suitable Fourier
expansion which allows to implement recursive estimates including the
true small divisors (\ref{5.trueres}). Nevertheless it is possible,
restricting to low and high frequencies, to improve the estimate of
Theorems \ref{3.alfateo} and \ref{3.betateo} with a modified PL
approach, thus discovering that $\beta$ QB thresholds are larger than
$\alpha$ ones.  Finally, we have also extended the one-parameter PL
perturbation scheme to a special $\alpha-\beta$ model, with
$\beta=\frac23\alpha^2$.  Thus the PL method developed, although
fitting the numerical simulations only for perturbations small enough,
gives a rigorous analytical way to describe localization in Fourier
space for any homogeneous FPU-like potential and any truncated Toda
potential.

We did not consider the linear stability fo QBs here. The reason is
that we focussed on the exponential localization of QBs. This property
we believe is not depending on whether the QB is stable or not. Let us
explain this point. If a QB is exponentially localized, the core can
not be in strong resonance with the tail modes - otherwise the QB
would not localize. Thus almost any instability which a QB may
encounter on the path of its continuation, will affect the dynamics of
the modes in the core of the excitation. Indeed, it has been observed
in \cite{DelLL95}, that such an instability triggers weak chaos among
the core modes, leaving the localization intact. The threshold to this
weak chaos state was correctly identified as an instability threshold
of QBs \cite{FlaIvaKan05}. Thus, given a QB is localized, its
stability or instability will not change the fact that nearby
trajectories (natural packets) will stay for long times in that phase
space part.

\newpage

%---------------------------------------------------------------------------
%---------------------------------------------------------------------------
%
%                     APPENDIX
%
%---------------------------------------------------------------------------
%---------------------------------------------------------------------------

\section{Appendix.}

\subsection{Appendix A.}

We give here a sketch of the proof of Prop. \ref{3.prop1}. We have
decided to omit the proofs of Prop. \ref{4.prop1} and
Prop. \ref{4.prop2} since, a part from small variations, they are
based on the same ideas.  

\smunp

\paragraph{Proof of Proposition \ref{3.prop1}:\,}

We give an inductive proof. For $k=2$ the statement is true. Suppose,
by inductive hypothesis, that it is also true until $k=n-1$. Let us
introduce here the following notation for the modes excited
$q_0=\overline{q}_1,\overline{q}_2,\ldots,\overline{q}_{n-1}$ and the
repeated dispersion law $F(x):=2|\sin(\frac{x\pi}{2N+2})|$; the
frequencies are respectively
\begin{displaymath}
F(q_0),\,F(2q_0),\,\ldots,F((n-1)q_0).
\end{displaymath}
By the inductive hypothesis, any mode ${\overline q}_j$ is obtained
from $jq_0$ through
\begin{displaymath}
\overline{q}_j =
\begin{cases}
{
J(2N+2)-jq_0, \qquad\qquad {F'(jq_0)>0},\cr
jq_0-J(2N+2), \qquad\qquad {F'(jq_0)<0},\cr
}
\end{cases}
\end{displaymath}
for some $J$ related to $jq_0$. From (\ref{3.eqTau2}) we know that the
equations of motion at the next order include products like
\begin{eqnarray*}
&B_{q {\overline q}_l {\overline q}_m}Q_{{\overline q}_l}^{(0)}Q_{{\overline q}_m}^{(n-2)}&\\
&B_{q {\overline q}_l {\overline q}_m}Q_{{\overline q}_l}^{(1)}Q_{{\overline q}_m}^{(n-3)}&\\
&\vdots&
\end{eqnarray*}
and, among them, the only ones giving a new frequency are
\begin{eqnarray*}
&Q_{\overline q_1}^{(0)}Q_{\overline q_{n-1}}^{(n-2)}&\\
&Q_{\overline q_2}^{(1)}Q_{\overline q_{n-2}}^{(n-3)}&\\
&\vdots&
\end{eqnarray*}
Indeed, by induction $Q_{{\overline q}_m}^{(n-h-1)}=0$ for $m>n-h$,
and $ q\pm \overline q_l\pm \overline q_m = 0 \,{\rm mod}(2N+2)$ is a
necessary condition for $B_{q {\overline q}_l {\overline
q}_m}\not=0$. We claim that the equations giving the new $q$ modes
\begin{eqnarray*}
&q \pm \overline{q}_1 \pm \overline{q}_{n-1}=0,2N+2,& \\
&q \pm \overline{q}_2 \pm \overline{q}_{n-2}=0,2N+2,&\\
&\vdots&
\end{eqnarray*}
all admit a common solution, which reads
\begin{displaymath}
\overline{q}_n =
\begin{cases}
{
\overline N(2N+2)-nq_0, \qquad\qquad {F'(nq_0)>0},\cr
nq_0-\overline N(2N+2), \qquad\qquad {F'(nq_0)<0}.\cr
}
\end{cases}
\end{displaymath}
As an example we consider just the second equation in the previous list. Let us
assume that
\begin{eqnarray*}
\overline{q}_2 &=& l(2N+2)-2q_0, \\
\overline{q}_{n-2} &=& L(2N+2)-(n-2)q_0,
\end{eqnarray*}
then the possible combinations are
\begin{eqnarray*}
\overline{q}_2 + \overline{q}_{n-2} &=& (L+l)(2N+2) - nq_0, \\
-\overline{q}_2 - \overline{q}_{n-2} &=& -(L+l)(2N+2) + nq_0, \\
\overline{q}_2 - \overline{q}_{n-2} &=& (l-L)(2N+2) + nq_0 - 4q_0, \\
-\overline{q}_2 + \overline{q}_{n-2} &=& (L-l)(2N+2) - nq_0 + 4q_0. 
\end{eqnarray*}
The last couple give the already included mode $\overline q_{n-4}$, so
we focus on the first couple, giving the four equations
\begin{eqnarray*}
q + \overline{q}_2 + \overline{q}_{n-2} &=& 0, \qquad\qquad 
q - \overline{q}_2 - \overline{q}_{n-2} = 0, \\
q + \overline{q}_2 + \overline{q}_{n-2} &=& 2N+2, \qquad
q - \overline{q}_2 - \overline{q}_{n-2} = 2N+2; \\
\end{eqnarray*}
notice that the first and the last above equations don't admit a solution $q\in[1,N]$, so
the only solutions are
\begin{eqnarray*}
q &=& 2N+2 - \overline{q}_2 - \overline{q}_{n-2} = nq_0 -
(L+l-1)(2N+2),\\ q &=& \overline{q}_2 + \overline{q}_{n-2} =
(L+l)(2N+2) - nq_0, \\
\end{eqnarray*}
when respectively $\overline{q}_2 + \overline{q}_{n-2}>N+1$ and
$\overline{q}_2 + \overline{q}_{n-2}<N+1$; they both correspond to the
same frequency $2|\sin{(\frac{nq_0\pi}{2N+2})}|$.

\begin{flushright}
$\blacksquare$
\end{flushright}

\subsection{Appendix B.}

Let 
\begin{eqnarray*}
R &:& (Q_j,P_j) \mapsto (Q_{j+1},P_{j+1})\cr
S &:& (Q_j,P_j) \mapsto (-Q_{2N+2-j},-P_{2N+2-j})\cr
\end{eqnarray*}
the discrete symmetries of the FPU system and, for a generic symmetry
$G$ on a manifold $\mathcal{M}$,
\begin{displaymath}
Fix(G) := \{ x \in \mathcal{M} \,|\, G(x)=x\}.
\end{displaymath}
It is possible to prove the following

\begin{proposition}
\label{3.prop2}
Let $q_0$ be a mode to be continued, $g_0:=gcd(2N+2,q_0)$ and
$\kappa_0:=\frac{2N+2}{g_0}$. Then, the $q_0$-breather sequence of
modes $\{\overline q_k\}$ coincides with $Fix(R^{\kappa_0})\cap
Fix(S)$ considered as an invariant submanifold of $Fix(S)$.
\end{proposition}
\begin{proof}
Firstly, following \cite{RINKinv}, we recall that 
\begin{eqnarray*}
  Fix(R^{\kappa_0}) &=& \{Q_j=P_j=0\,\, {\rm iff} j\not= 0\, {\rm
  mod}(g_0)\},\\
  Fix(S) &=& \{Q_j = -Q_{2N+2-j},\,P_j = -P_{2N+2-j}\},\\
\end{eqnarray*}
so that it is possible to interpret $Fix(S)$ as the embedding of the
fixed boundary condition system in the periodic boundary condition
system. The thesis reduces to
\begin{displaymath}
\{j= 0\, {\rm mod}(g_0)\} = \{\overline q_k\},\qquad\qquad j=1,\ldots,N.
\end{displaymath}
In order to show the double inclusion of the two sets it is useful to
remind that, from the definition of $g_0$, it follows 
\begin{equation}
\alpha q_0+\beta (2N+2) = \pm 1\qquad\qquad \alpha,\beta \in\mathbb{Z}.
\end{equation}
\begin{description}

\item[$\subseteq$)] let us take $j = 0\,{\rm mod}(g_0)$, which means
  $j=Jg_0$ with $J\leq [\frac{N+1}{g_0}]$; we have that
\begin{displaymath} 
Aq_0 + B(2N+2) = \pm Jg_0,\qquad A=\alpha J,\,B=\beta J.
\end{displaymath} 
\item[$\supseteq$)] this case is trivial, since the definition in
  Proposition \ref{3.prop1} implies $g_0/\overline q_k$.

\end{description}
\begin{flushright}
$\blacksquare$
\end{flushright}
\end{proof}

Proposition \ref{3.prop2} claims that we should interpret the fixed
boundary conditions system as embedded in the periodic one and then
consider the symmetries $Fix(R^\kappa)$ with $2(N+1)=\kappa g$;
observe that it must be $\kappa\not=2$, otherwise $q_0=N+1>N$. So, any
QB is a special orbit of one of such submanifolds. Observe also that,
for $\kappa=2N+2$ the invariant submanifold coincides with the fixed
boundary subsystem $Fix(R^{\kappa})\cap Fix(S)=Fix(S)$, while for
$\kappa=2$ we have $Fix(R^{\kappa})\cap Fix(S)=\{0\}$. All the others
submanifolds $Fix(R^{\kappa})\cap Fix(S)$ with $\kappa$ not being a
divisor of $2N+2$ are meaningless since

\begin{proposition}
\label{3.prop3}
For all $l=1,\ldots,2N+2$ holds true
\begin{equation}
Fix(R^{l}) = Fix(R^{{\rm gcd}(l,2N+2)}).
\end{equation}
\end{proposition}
\begin{proof}
Simply, by calling $\kappa={\rm gcd}(l,2N+2)$ and $2N+2 = \kappa g$
\begin{displaymath}
\frac{2N+2}{{\rm gcd}(l,2N+2)} = g = \frac{2N+2}{{\rm gcd}(\kappa,2N+2)}.
\end{displaymath}

\begin{flushright}
$\blacksquare$
\end{flushright}
\end{proof}

The final scenario is the following. Take $1\leq q_0\leq N$ and
$g_0,\,\kappa_0$ ad defined in the Proposition \ref{3.prop2}, then

\begin{enumerate}
 \item $g_0=1\,\Leftrightarrow \kappa_0=2N+2\Leftrightarrow
 Fix(R^{\kappa_0})\cap Fix(S)=Fix(S)$: the submanifold is trivial.

 \item $g_0\geq 2$ and $\frac{2N+2}{g_0}\not\in\mathbb{N} \Rightarrow$
 the submanifold doesn't coincide with $Fix(S)$ but cannot represent
 a fixed boundary conditions subsystem. Hence, the QB does not involve
 all the modes (not trivial), but it is not a rescaled solution of a
 lower dimensional system.

 \item $g_0\geq 2$ and $g_0=gcd(N+1,q_0) \Rightarrow$ the submanifold
 is the embedding of a fixed boundary condition subsystem with
 $\frac{N+1}{g_0}$ particles and the QB is the $\frac{q_0}{g_0}$-mode
 rescaled solution, constructed as in \cite{KanFlaIvaMsh06}.
\end{enumerate}

\subsection{Appendix C.}

It is known (cfr. \cite{GM2002}) that the sequence
\begin{equation}
\label{A.seq}
\begin{cases}
{
\lambda_1 = 1,\cr
\lambda_n = \sum_{j=1}^{n-1}{\lambda_j\lambda_{n-j}} = \sum_{i+j=n}{\lambda_i\lambda_j}\cr
}
\end{cases}
\end{equation}
is equivalent to a suitable functional equation
\begin{equation}
\label{A.functeq}
g(z) = z + g^2(z),\qquad\qquad g(z) = \sum_{r\geq 1}{\lambda_rz^r}
\end{equation}
and that follows $\lambda_r < 4^{r-1}$. Consider now the modified
problem
\begin{equation}
\begin{cases}
{
\lambda_1 = \mu_1,\cr
\lambda_n = \mu_n\sum_{j=1}^{n-1}{\lambda_j\lambda_{n-j}}\cr
}
\end{cases},
\end{equation}
where $\mu_j$ is a limited sequence. Let $\mu:=\sup_j{\mu_j}$ and
define the auxiliary sequence
\begin{equation}
\begin{cases}
{
\Lambda_1 = \mu,\cr
\Lambda_n = \mu\sum_{j=1}^{n-1}{\Lambda_j\Lambda_{n-j}}\cr
}
\end{cases},
\end{equation}
whose related function satisfies
\begin{equation}
g(z) = \mu z + \mu g^2(z).
\end{equation}
It is easy to prove by induction that
\begin{displaymath}
\lambda_n = \mu_n\sum_{j=1}^{n-1}{\lambda_j\lambda_{n-j}}\leq
\mu\sum_{j=1}^{n-1}{\Lambda_j\Lambda_{n-j}} = \Lambda_n.
\end{displaymath}
Thus, a bound for $\lambda_n$ can be obtained from the
following asymptotic behaviour of $\Lambda_n$
\begin{displaymath}
\Lambda_n = \mu (2\mu^2)^{n-1}\frac{(2n-3)!!}{n!}\sim
\frac{\mu}{n}(4\mu^2)^{n-1}.
\end{displaymath}
This result can be applied to Theorem \ref{3.alfateo}, where we deal
with the sequence
\begin{displaymath}
c_n := \omega_{nq_0}\sum_{l=1}^{n-1}{\omega_{lq_0}\omega_{(n-l)q_0}c_l c_{n-l}},
\end{displaymath}
which reads also
\begin{displaymath}
\lambda_n := \omega_{nq_0}^2\sum_{l=1}^{n-1}{\lambda_{l}\lambda_{n-l}},
\end{displaymath}
where $\lambda_l=\omega_{lq_0}c_l$ and $\omega_l^2<4$.

Let us move to a more complex sequence
\begin{equation}
\label{2.seq}
\begin{cases}
{
\lambda_1 = 1,\cr
\lambda_n = \sum_{i+j+k=n+1}{\lambda_i\lambda_j\lambda_k}\cr
}
\end{cases},
\end{equation}
where plainly $1\leq i,j,k \leq n-1$. Following \cite{MR990993}, we
define an analytic function $g(z)=\sum_{r\geq 1}{\lambda_rz^r}$ such
that (\ref{2.seq}) is equivalent to the functional equation
\begin{equation}
\label{2.functeq}
g = z + \frac1z g^3;
\end{equation}
we stress that, since $g$ has a first order zero in $z=0$, then also
$\frac1z g(z)$ is analytic. Let us define an auxiliary function $f(z) =
\frac1z g(z) = \sum_{r\geq 0}{c_rz^r}$, with 
$c_r=\lambda_{r+1}$, and rewrite (\ref{2.functeq}) as
\begin{equation}
\label{2.newfuncteq}
f = 1 + z f^3,\qquad\qquad f(0) = c_0 = \lambda_1 = 1.
\end{equation}
The differentiation of (\ref{2.newfuncteq}) gives
\begin{displaymath}
f' = \frac{f^3}{1-3zf^2},
\end{displaymath}
which, exploiting again (\ref{2.newfuncteq}), reads as a $z$-independent
expression
\begin{equation}
\label{2.diffeq}
f'=\frac{f^4}{3-2f}.
\end{equation}
We may assume the following general description for $f^{(r-1)}$
\begin{equation}
\label{2.guess}
f^{(r-1)} = \sum_{j=0}^{2r-3}{\sum_{k=0}^{4r}{\beta_{jk}\frac{f^k}{(3-2f)^j}}},
\end{equation}
with suitable values for $\beta_{jk}$, and check its reliability by
induction. The differentiation of the generic addendum in (\ref{2.guess})
gives
\begin{displaymath}
\frac d {dz} \frac{f^k}{(3-2f)^j} = \frac{kf^{k+3}}{(3-2f)^{j+1}} + \frac{2jf^{k+4}}{(3-2f)^{j+2}},
\end{displaymath}
which yields
\begin{equation}
f^{(r)} = \sum_{j=0}^{2r-1}{\sum_{k=0}^{4r+4}{\alpha_{jk}\frac{f^k}{(3-2f)^j}}},
\end{equation}
where the new coefficients $\alpha_{jk}$ are properly defined through
$\beta_{jk}$. As noticed in \cite{MR990993}, Lemma 11.2, we can avoid the
computations of $\alpha_{jk}$ simply using the previous computation
\begin{eqnarray*}
f^{(r)}(0) &=& \sum_{j=0}^{2r-3}{\sum_{k=0}^{4r}{
\beta_{jk}\frac{kf(0)^{k+3}}{(3-2f(0))^{j+1}} +
\frac{2jf(0)^{k+4}}{(3-2f(0))^{j+2}} }} = \cr &=&
\sum_{j=0}^{2r-3}{\sum_{k=0}^{4r}{ \beta_{jk} (k+2j)}},
\end{eqnarray*}
which allows to find
\begin{displaymath}
c_r = \frac1{r!}f^{(r)}(0) < \frac1{r!}8r \sum_{j=0}^{2r-3}{\sum_{k=0}^{4r}{
    \beta_{jk}}} = 8r \frac1{r!} f^{(r-1)}(0) = 8 c_{r-1}.
\end{displaymath}
This concludes the proof, since we have finally shown
\begin{displaymath}
\lambda_n\leq 8^{n-1}.
\end{displaymath}

Let us consider the sequence
\begin{equation}
\label{3.seq}
\begin{cases}
{
\mu_1 = 1,\cr
\mu_2 = 1,\cr
\mu_n = \sum_{i+j+k=n}{\mu_i\mu_j\mu_k}\cr
}
\end{cases}.
\end{equation}
Here the solution comes from the comparison of (\ref{3.seq}) with the
previous sequence (\ref{2.seq}). Indeed, with the choice
$\lambda_1=\mu_1=\mu_2=1$, it is possible to prove by induction that
$\mu_n\leq\lambda_{n-1}\leq\lambda_n.$

%---------------------------------------------------------------------------
%---------------------------------------------------------------------------
%
%                     BIBLIOGRAPHY
%
%---------------------------------------------------------------------------
%---------------------------------------------------------------------------

\newpage
%BIBLIOGRAFIA

\vfill
\newpage

\section*{List of captions}
\begin{enumerate}

\item {\bf FIG.1}: {(a): Natural packet evolution initially exciting mode
$q_0=1$. The energies of normal modes are plotted versus $q$. Circles
- $10^4$, squares - $10^5$, rhombs - $10^6$.  Dashed line -
$q$-breather from (b) for comparison.  (b): The energies of normal
modes versus $q$ for a $q$-breather with $q_0=1$.  (c): sequence
$\eta_j^{-1}$.  (d): sequence $\chi_j^{-1}$.  Parameters in all cases:
$N=31,\,\alpha=0.33,\,\mathcal{E}=0.01$.}

\item {\bf FIG.2}: { (a): Natural packet evolution initially exciting mode
$q_0=2$ (circles).  The energies of normal modes are plotted versus
$q$ at $t=10^5$.  The dashed line shows the energy distribution for a
corresponding $q$-breather with $q_0=2$.  (b): Same as (a), but for
$q_0=3$.  (c): sequence $\eta_j^{-1}(2,47)$.  (d): sequence
$\eta_j^{-1}(3,47)$.  Parameters in all cases:
$N=47,\,\alpha=0.1,\,\mathcal{E}=0.0025$.}

\item {\bf FIG.3}: {Parameters:
$N=47,\,\alpha=0.1,\,\mathcal{E}=0.0025$. (a): Energy of normal modes
versus mode number of a QB for $q_0=2$. (b): Similar to (a), but for a
QB with $q_0=3$. (c): sequence $\chi_j^{-1}(2,47)$. (d): sequence
$\chi_j^{-1}(3,47)$.}

\item {\bf FIG.4}: {Parameters: $N=47,\,\mathcal{E}=0.1$. (a): Energy
of normal modes in a QB (open symbols) and a natural packet (filled
symbols, waiting time $10^5$) for $q_0=47$ and $\alpha=0.25$. (b):
Similar to (a), but for $\beta=0.25$. (c): sequence $\chi_j^{-1}$
which corresponds to case (a).  (d): sequence $\chi_j^{-1}$ which
corresponds to (b).}

\end{enumerate}

\end{document}